\documentclass[pre, twocolumn, showkeys, amsmath, amssymb]{revtex4}

\usepackage{graphicx}

\begin{document}

\title{Semiconservative quasispecies equations for polysomic genomes:  The general case}

\author{Eran Itan}
\affiliation{Department of Chemistry, Ben-Gurion University of the Negev, Be'er-Sheva, Israel}
\author{Emmanuel Tannenbaum}
\email{emanuelt@bgu.ac.il}
\affiliation{Department of Chemistry, Ben-Gurion University of the Negev, Be'er-Sheva, Israel}

\begin{abstract}

This paper develops a formulation of the quasispecies equations appropriate for polysomic, semiconservatively replicating genomes.  This paper is an extension
of previous work on the subject, which considered the case of haploid genomes.  Here, we develop a more general formulation of the quasispecies equations that
is applicable to diploid and even polyploid genomes.  Interestingly, with an appropriate classification of population fractions, we obtain a system of
equations that is formally identical to the haploid case.  As with the work for haploid genomes, we consider both random and immortal DNA strand 
chromosome segregation mechanisms.  However, in contrast to the haploid case, we have found that an analytical solution for the mean fitness is considerably more difficult to obtain for the polyploid case.  Accordingly, whereas for the haploid case we obtained expressions for the mean fitness for the case of an analogue of the single-fitness-peak landscape for arbitrary lesion repair probabilities (thereby allowing for non-complementary genomes), here we solve for the mean fitness for the restricted case of perfect lesion repair.

\end{abstract}


\maketitle
                                                  
\section{Introduction}

The quasispecies theory of evolutionary dynamics was originally introduced in a now-classic paper by Manfred Eigen in 1971 \cite{Eigen}.  In this paper, Eigen developed a system of ordinary differential equations that were meant to describe the evolutionary dynamics of replicating polynucleotide or polypeptide chains.  The goal was to develop a mathematical framework that would be suitable for modeling the evolutionary processes relevant to the origin of life.  Much of the subsequent work by Eigen on quasispecies theory was done in collaboration with Peter Schuster, which is why the quasispecies equations are often referred to as the {\it Eigen-Schuster equations} \cite{EigenSchuster}.

In brief, the quasispecies model considers a population of genomes, defined as single-stranded sequences, taken to be of length $ L $.  A given sequence, denoted $ \sigma $, may be expressed as $ \sigma = s_1 s_2 \dots s_L $, where each $ s_i $ represents a ``letter" or ``base" that is chosen from an alphabet of size $ S $ (for all known terrestrial life, $ S = 4 $, though many phenomenological studies work with $ S = 2 $ for simplicity) \cite{EigenSchuster, TannQuasReview, WilkeQuasReview, BullQuasReview}.

With each $ \sigma $ is associated a first-order growth rate constant, denoted by $ \kappa_{\sigma} $.  The mapping $ K: \sigma \rightarrow \kappa_{\sigma} $ defines what is known as the {\it fitness landscape}.  During replication, it is assumed that a daughter strand is produced from the template parent strand.  Replication is not necessarily error-free, which gives rise to a transition probability $ p_m(\sigma, \sigma') $, denoting the probability that parent strand $ \sigma $ produces the daughter $ \sigma' $.  The quasispecies equations may then be expressed as \cite{EigenSchuster, TannQuasReview, WilkeQuasReview, BullQuasReview},
\begin{equation}
\frac{d x_{\sigma}}{dt} = \sum_{\sigma'} \kappa_{\sigma'} p_m(\sigma', \sigma) x_{\sigma'} - \bar{\kappa}(t) x_{\sigma} 
\end{equation}
Here, $ x_{\sigma} $ denotes the fraction of organisms in the population that have genome $ \sigma $, and $ \bar{\kappa}(t) \equiv \sum_{\sigma} \kappa_{\sigma} x_{\sigma} $ is the {\it mean fitness} of the population.

The central result of quasispecies theory is a phenomenon known as the {\it error catastrophe}.  The error catastrophe refers to a localization to de-localization transition over the genome sequence space once mutation rates have crossed a critical threshold, naturally termed the {\it error threshold}.  Below the error threshold, natural selection is sufficiently strong to localize the population distribution to a ``cloud" of related strains, termed a quasispecies.  Above the error threshold, natural selection is no longer able to counteract mutation-accumulation, and the result is evolutionary dynamics governed by essentially random genetic drift.  Over time, the population distribution completely de-localizes over the sequence space, and no identifiable quasispecies emerges.

Although the origin-of-life problem was the original motivation for the development of quasispecies theory, the quasispecies concept has found broad application in the field of virus evolutionary dynamics.  The reason for this is that many RNA viruses, such as HIV, have sufficiently high mutation rates that they exhibit a fairly broad distribution of genotypes, so that the quasispecies concept is highly relevant for these systems.  However, because the quasispecies equations may be readily adapted toward modeling evolution in more complex systems, in recent years there have been efforts to develop quasispecies theory into a useful framework for modeling the evolution of cell-based life.  Understanding evolution at the cellular level will have applications in areas such as antibiotic drug resistance in bacteria, immune system function, stem cells, and the somatic evolution of cancer.

Some of the work that has been done in quasispecies theory to make it suitable for modeling biological systems more complex than molecules and viruses includes the following:  (1)  Developing a formulation of the quasispecies model that is appropriate for double-stranded, semiconservatively replicating DNA genomes \cite{TannSemiConserve}.  (2)  Analysis of quasispecies dynamics for multi-gene genomes, which, among other results, revealed that the error catastrophe is a special case of a more general phenomenon that was termed an ``error cascade" \cite{TannMultiGene}.  (3)  Using quasispecies theory to model evolution in dynamic environments, and to study the co-evolutionary dynamics that arises from the immune response to a viral infection \cite{NilSnoadDyn, KampBornholdtBCell}.  (4)  Modeling mutation-propagation in stem and tissue cells \cite{TannStemCell}.  (5)  Modeling genetic repair and repair-deficient strains known as mutators \cite{KesslerRepair, NowakRepair, TannRepair}.  (6)  Incorporating Horizontal Gene Transfer and recombination into quasispecies theory \cite{KesslerHGT, DeemHGT1, DeemHGT2}.

Additionally, other recent work on quasispecies theory has included developing quasispecies equations appropriate for describing polysomic genomes \cite{TannHaploid}.  Given that cellular genomes are generally composed of several chromosomes, such a formulation of the quasispecies model is a necessary extension for developing realistic models of the evolutionary dynamics of cellular populations.  However, the work on polysomic genomes only considered haploid genomes.  Here, in this work, we generalize the quasispecies equations for polysomic genomes to allow for polyploid genomes.  We do not use our equations to model a specific biological system in this paper.  Nevertheless, we obtain analytical results for the polysomic analogue of the single-fitness-peak landscape, which is the simplest and most commonly studied fitness landscape in quasispecies theory.  These analytical results are in agreement with results obtained from stochastic simulations, suggesting that the equations developed here may be suitable for modeling evolutionary processes in real systems.  

\section{The Model}

\subsection{The Finite Sequence Length Equations}

We consider a population of asexually replicating organisms, each of which is characterized by a genome consisting of $ N $ chromosomes.  Unlike our previous paper \cite{HAPLOID}, we do not assume that the chromosomes are necessarily distinguishable, so that we do not impose any kind of chromosome ordering.  Thus, a given genome, denoted $ \hat{\sigma} $, may be written as $ \hat{\sigma} = \{\{\sigma_1, \sigma_1'\}, \dots, \{\sigma_N, \sigma_N'\}\} $, where $ \{\sigma_i, \sigma_i'\} $ denotes the pair of DNA strands of the $ i^{\mbox{th}} $ chromosome.  We also assume that the organisms replicate at a rate characterized by
a genome-dependent first-order growth rate constant $ \kappa_{\hat{\sigma}} $.  

Furthermore, we let $ p((\sigma''; \hat{\sigma}''), \{\sigma, \sigma'\}) $ denote the probability that strand $ \sigma'' $, as part of genome $ \hat{\sigma}'' $, becomes, after daughter strand synthesis and post-replication lesion repair, chromosome $ \{\sigma, \sigma'\} $.  We also let $ p((\sigma''; \hat{\sigma}''), (\sigma, \sigma')) $
denote the probability that strand $ \sigma'' $, as part of genome $ \hat{\sigma}'' $, becomes, after daughter strand synthesis and post-replication lesion repair, strand $ \sigma $, with daughter strand $ \sigma' $.  It should be noted that,
\begin{widetext}
\begin{equation}
p((\sigma''; \hat{\sigma}''), \{\sigma, \sigma'\}) =
\left\{ 
\begin{array}{cc}
p((\sigma''; \hat{\sigma}''), (\sigma, \sigma')) + p((\sigma''; \hat{\sigma}''), (\sigma', \sigma)) 
&
\mbox{ if $ \sigma \neq \sigma' $} \\
p((\sigma''; \hat{\sigma}''), (\sigma, \sigma'))
&
\mbox{ if $ \sigma = \sigma' $}\\
\end{array}
\right.
\end{equation}
\end{widetext}

\subsubsection{Random chromosome segregation}

We first consider the case of random chromosome segregation.  Given a population of replicating organisms, we let $ x_{\hat{\sigma}} $ denote the fraction of
the population characterized by the genome $ \hat{\sigma} $.  Our goal is to develop an expression for $ d x_{\hat{\sigma}}/dt $.  To do so, we note
that the expression for $ d x_{\hat{\sigma}}/dt $ consists of three separate terms:  (1) A destruction term, corrresponding to the effective destruction
of the parent genome as a result of semiconservative replication \cite{HAPLOID}. (2) A mean-fitness normalization term, that arises when converting the dynamical
equations expressed in terms of population numbers into dynamical equations expressed in terms of population fractions.  (3) A mutation contribution term, summing the contribution to $ x_{\hat{\sigma}} $ from the various genomes in the population.  From Appendix A, we have that,
\begin{widetext}
\begin{eqnarray}
\frac{d x_{\hat{\sigma}}}{dt} 
& = &
-(\bar{\kappa}(t) + \kappa_{\hat{\sigma}}) x_{\hat{\sigma}}
+ \frac{1}{2^{N-1}} \sum_{\hat{\sigma}'' = \{\{\sigma_1'', \sigma_1'''\}, \dots, \{\sigma_N'', \sigma_N'''\}\}}
\kappa_{\hat{\sigma}''} x_{\hat{\sigma}''} \times
\nonumber \\
&   &
\sum_{\pi_N \in \pi_N(\hat{\sigma})} 
\prod_{i = 1}^{N}
[p((\sigma_i''; \hat{\sigma}''), \{\sigma_{\pi_N(i)}, \sigma_{\pi_N(i)}'\}) +
 p((\sigma_i'''; \hat{\sigma}''), \{\sigma_{\pi_N(i)}, \sigma_{\pi_N(i)}'\})] 
\end{eqnarray}
\end{widetext}
where $ \pi_N $ denotes a permutation of the indices $ 1, \dots, N $, and $ \pi_N(\hat{\sigma}) $ denotes the subset of all such permutations that gives rise to distinct vectors of strand-pairs $ (\{\sigma_{\pi_N(1)}, \sigma_{\pi_N(1)}'\}, \dots, \{\sigma_{\pi_N(N)}, \sigma_{\pi_N(N)}'\}) $ obtained from the genome
$ \hat{\sigma} = \{\{\sigma_1, \sigma_1'\}, \dots, \{\sigma_N, \sigma_N'\}\} $.

We may switch from an unordered chromosome representation of the genome, to an ordered one, as follows:  Given a genome $ \{\{\sigma_1, \sigma_1'\}, \dots,
\{\sigma_N, \sigma_N'\}\} $, let $ m $ denote the number of distinct strand-pairs.  Then we may write that this genome consists of the $ m $ distinct strand-pairs $ \{\sigma_{i_1}, \sigma_{i_1}'\}, \dots, \{\sigma_{i_m}, \sigma_{i_m}'\} $, where the strand-pair $ \{\sigma_{i_k}, \sigma_{i_k}'\} $ appears $ n_k $ times, so that $ n_1 + \dots + n_m = N $.

Note that there are $ N!/(n_1! \times \dots \times n_m!) $ distinct permutations of $ (\{\sigma_1, \sigma_1'\}, \dots, \{\sigma_N, \sigma_N'\}) $, so define,
\begin{eqnarray}
&    &
x_{\tilde{\sigma} = (\{\sigma_1, \sigma_1'\}, \dots, \{\sigma_N, \sigma_N'\})} = 
\nonumber \\
&    &
\frac{n_1! \times \dots \times n_m!}{N!} 
x_{\{\{\sigma_1, \sigma_1'\}, \dots, \{\sigma_N, \sigma_N'\}\}}
\end{eqnarray}

We obtain, again following the derivation provided in Appendix A,
\begin{eqnarray}
\frac{d x_{\tilde{\sigma}}}{dt} 
& = &
-(\bar{\kappa}(t) + \kappa_{\tilde{\sigma}}) x_{\tilde{\sigma}}
+ \frac{1}{2^{N-1}}
\sum_{\tilde{\sigma}''} \kappa_{\tilde{\sigma}''} x_{\tilde{\sigma}''}
\times \nonumber \\
&   &
\prod_{i = 1}^{N} 
[p((\sigma_i''; \tilde{\sigma}''), \{\sigma_i, \sigma_i'\}) +
 p((\sigma_i'''; \tilde{\sigma}''), \{\sigma_i, \sigma_i'\})]
\nonumber \\
\end{eqnarray}
where $ p((\sigma_i''; \tilde{\sigma}''), \{\sigma_i, \sigma_i'\}) $ denotes the probability that parent strand $ \sigma_i'' $, as part of genome $ \tilde{\sigma}'' $, becomes, after daughter strand synthesis and lesion repair, chromosome $ \{\sigma_i, \sigma_i'\} $.

Proceeding as with the case of haploid genomes, we may define a vector of ordered strand-pairs population fraction via the definition,
\begin{eqnarray}
x_{\vec{\sigma}} = x_{((\sigma_1, \sigma_1'), \dots, (\sigma_N, \sigma_N'))} = 
\frac{1}{2^k} x_{\tilde{\sigma}}
\end{eqnarray}
where $ k $ denotes the number of chromosomes for which $ \sigma_i \neq \sigma_i' $.  As with the case for haploid genomes, we obtain from Appendix A that,
\begin{eqnarray}
\frac{d x_{\vec{\sigma}}}{dt}
& = &
-(\bar{\kappa}(t) + \kappa_{\vec{\sigma}}) x_{\vec{\sigma}}
+ \frac{1}{2^{N-1}}
\sum_{\vec{\sigma}''} \kappa_{\vec{\sigma}''} x_{\vec{\sigma}''}
\times \nonumber \\
&   &
\prod_{i = 1}^{N}
[p((\sigma_i''; \vec{\sigma}''), (\sigma_i, \sigma_i')) +
 p((\sigma_i''; \vec{\sigma}''), (\sigma_i', \sigma_i))]
\nonumber \\
\end{eqnarray}

\subsubsection{Immortal DNA strand co-segregation}

Immortal DNA strand co-segregation is a chromosome segregation mechanism whereby one of the daughter cells receives all of the chromosomes containing the oldest DNA template strands of the previous replication cycle.  It is a chromosome segregation mechanism that was hypothesized to be at work in adult stem cells \cite{Cairns}, as a way to reduce the accumulation of mutations in stem cells.  Immortal DNA strand co-segregation has been experimentally confirmed \cite{Potten, Sherley}.  Interestingly, there is evidence to suggest that even unicellular organisms, such as {\it Saccharomyces cerevisiae}, may exhibit immortal DNA strand co-segregation \cite{ThorpeYeastImmortal}.  As a result, we will develop the quasispecies equations for immortal strand segregation as well.

To derive the equations for immortal DNA strand co-segregation, we first note that a given DNA strand in a genome is either newly synthesized, or it
has gone through a previous replication cycle where it was a template strand.  Once a DNA strand is a template strand, then it remains a template strand throughout all successive replications.  Given a strand $ \sigma $, we let $ \sigma^{(N)} $ denote a strand that is ``new,'' that is, it has never been a template strand, and we let $ \sigma^{(T)} $ denote a strand that has been a template strand at least once.  Since a chromosome that was produced in a replication cycle must consist of exactly one template and one new strand, a given chromosome is either of the form $ \{\sigma^{(N)}, \sigma'^{(N)}\} $, or $ \{\sigma^{(T)}, \sigma'^{(N)}\} $.

We also note that a given genome consists entirely of chromosomes containing only new strands, or entirely of chromosomes containing one template and one
new strand.  For if one chromosome contains a template strand, then that strand must have come from a parent cell in a previous replication cycle.  This parent 
cell must have had $ N - 1 $ other parent strands coming from $ N - 1 $ other chromosomes that segregated into the daughter cell.  Therefore, the other chromosomes of the genome must contain a template strand as well.  

Given a genome $ \hat{\sigma} $, we let $ \hat{\sigma}^{(N/N)} $ signify that the genome consists entirely of new strands, and we let $ \hat{\sigma}^{(T/N)} $
signify that the genome consists of chromosomes containing exactly one template and one new strand.  We then have, from Appendix A,
\begin{widetext}
\begin{eqnarray}
&   &
\frac{d x_{\hat{\sigma}^{(N/N)}}}{dt} =
-(\bar{\kappa}(t) + \kappa_{\hat{\sigma}}) x_{\hat{\sigma}^{(N/N)}}
\nonumber \\
&   &
\frac{d x_{\hat{\sigma}^{(T/N)}}}{dt} =
-(\bar{\kappa}(t) + \kappa_{\hat{\sigma}}) x_{\hat{\sigma}^{(T/N)}}
+ \frac{1}{2^{N-1}} 
\sum_{\hat{\sigma}''^{(N/N)} = \{\{\sigma_1''^{(N)}, \sigma_1'''^{(N)}\}, \dots, \{\sigma_N''^{(N)}, \sigma_N'''^{(N)}\}\}}
\kappa_{\hat{\sigma}''} x_{\hat{\sigma}''^{(N/N)}}
\times 
\nonumber \\
&   &
\sum_{\pi_N \in \pi_N(\hat{\sigma}^{(T/N)}}
\prod_{i = 1}^{N}
[p((\sigma_i''; \hat{\sigma}''), (\sigma_{\pi_N(i)}, \sigma_{\pi_N(i)}')) +
 p((\sigma_i'''; \hat{\sigma}''), (\sigma_{\pi_N(i)}, \sigma_{\pi_N(i)}'))]
\nonumber \\
&   &
+ 
\nonumber \\
&   &
\sum_{\hat{\sigma}''^{(T/N)} = \{\{\sigma_1''^{(T)}, \sigma_1'''^{(N)}\}, \dots, \{\sigma_N''^{(T)}, \sigma_N'''^{(N)}\}\}}
\kappa_{\hat{\sigma}''} x_{\hat{\sigma}''^{(T/N)}}
\times 
\nonumber \\
&   & 
\sum_{\pi_N \in \pi_N(\hat{\sigma}^{(T/N)}}
[\prod_{i = 1}^{N} p((\sigma_i''; \hat{\sigma}''), (\sigma_{\pi_N(i)}, \sigma_{\pi_N(i)}')) + 
 \prod_{i = 1}^{N} p((\sigma_i'''; \hat{\sigma}''), (\sigma_{\pi_N(i)}, \sigma_{\pi_N(i)}'))]
\end{eqnarray}
\end{widetext}

As with the random segregation equations, we can define an ordered chromosome formulation of the dynamics for the immortal DNA strand equations.  We obtain, from Appendix A,
\begin{widetext}
\begin{eqnarray}
&   &
\frac{d x_{\tilde{\sigma}^{(N/N)}}}{dt} = 
-(\bar{\kappa}(t) + \kappa_{\tilde{\sigma}}) x_{\tilde{\sigma}^{(N/N)}}
\nonumber \\
&   &
\frac{d x_{\tilde{\sigma}^{(T/N)}}}{dt} =
-(\bar{\kappa}(t) + \kappa_{\tilde{\sigma}}) x_{\tilde{\sigma}^{(T/N)}}
+ \frac{1}{2^{N-1}}
\times
\nonumber \\
&   &
\sum_{\tilde{\sigma}''^{(N/N)}}
\kappa_{\tilde{\sigma}''} x_{\tilde{\sigma}''^{(N/N)}}
\prod_{i = 1}^{N} 
[p((\sigma_i''; \tilde{\sigma}''), (\sigma_i, \sigma_i')) + p((\sigma_i'''; \tilde{\sigma}''), (\sigma_i, \sigma_i'))]
\nonumber \\
&   &
+ \sum_{\tilde{\sigma}''^{(T/N)}}
\kappa_{\tilde{\sigma}''} x_{\tilde{\sigma}''^{(T/N)}}
[\prod_{i = 1}^{N} p((\sigma_i''; \tilde{\sigma}''), (\sigma_i, \sigma_i')) + 
 \prod_{i = 1}^{N} p((\sigma_i'''; \tilde{\sigma}''), (\sigma_i, \sigma_i'))]
\nonumber \\
\end{eqnarray}
\end{widetext}

Now, define an ordered strand-pair formulation of the dynamics as follows:  Define
\begin{equation}
y_{\vec{\sigma}^{(N/N)}} = \frac{1}{2^k} x_{\tilde{\sigma}^{(N/N)}}
\end{equation}
and
\begin{equation}
y_{\vec{\sigma}^{(T/N)}} = x_{\tilde{\sigma}^{(T/N)}}
\end{equation}
and
\begin{equation}
y_{\vec{\sigma}} = y_{\vec{\sigma}^{(N/N)}} + y_{\vec{\sigma}^{(T/N)}}
\end{equation}
where $ k $ denotes the number of chromosomes with distinct strands
in the genome.  We then have, from Appendix A,
\begin{eqnarray}
\frac{d y_{\vec{\sigma}}}{dt} 
& = &
-(\bar{\kappa}(t) + \kappa_{\vec{\sigma}}) y_{\vec{\sigma}}
+ \sum_{\vec{\sigma}''} \kappa_{\vec{\sigma}''} y_{\vec{\sigma}''}
\times 
\nonumber \\
&   &
[\prod_{i = 1}^{N} p((\sigma_i''; \vec{\sigma}''), (\sigma_i, \sigma_i')) +
 \prod_{i = 1}^{N} p((\sigma_i'''; \vec{\sigma}''), (\sigma_i, \sigma_i'))]
\nonumber \\
\end{eqnarray}

\subsubsection{Complementarity symmetry}

It is interesting to note that even when we do not assume that the genomes are necessarily haploid, it still follows that it is possible to derive an ordered 
strand-pair formulation of the dynamics that is identical to the haploid case \cite{TannHaploid}.  We should therefore note that in the case of haploid genomes, we made an
additional assumption regarding the fitness and error landscapes that allows for a convenient representation of the dynamics.  We make the identical assumption in this paper, and obtain a similarly convenient representation of the dynamics for the general case.

Following \cite{HAPLOID}, we begin by defining two operations $ \tau $ and $ \gamma $, acting on ordered strand-pairs, as follows:  $ \tau (\sigma, \sigma') =
(\sigma', \sigma) $, and $ \gamma (\sigma, \sigma') = (\bar{\sigma}, \bar{\sigma}') $, where $ \bar{\sigma} $ denotes the strand complementary to $ \sigma $ (because DNA is antiparallel, then if $ \sigma = b_1 \dots b_L $, and if $ \bar{b}_i $ denotes the base complementary to $ b_i $, we then have $ \bar{\sigma} = \bar{b}_L \dots \bar{b}_1 $.  Furthermore, given some vector of ordered strand-pairs $ \vec{\sigma} = ((\sigma_1, \sigma_1'), \dots, (\sigma_N, \sigma_N')) $, and a vector $ \vec{s} = (s_1, \dots, s_N) $, with each $ s_i = 0, 1 $, we make the following definitions:
\begin{eqnarray}
&   &
\tau^{\vec{s}} \vec{\sigma} = (\tau^{s_1} (\sigma_1, \sigma_1'), \dots, \tau^{s_N} (\sigma_N, \sigma_N'))
\nonumber \\
&   &
\gamma^{\vec{s}} \vec{\sigma} = (\gamma^{s_1} (\sigma_1, \sigma_1'), \dots, \gamma^{s_N} (\sigma_N, \sigma_N'))
\end{eqnarray}

Now, note that the fitness landscape is symmetric under $ \tau $, that is $ \kappa_{\tau^{\vec{s}} \vec{\sigma}} = \kappa_{\vec{\sigma}} $ for all $ \vec{s} \in \{0, 1\}^N $.  We also assume that the fitness landscape satisfies a {\it complementarity symmetry}, that is, $ \kappa_{\gamma^{\vec{s}} \vec{\sigma}} = \kappa_{\vec{\sigma}} $ for all $ \vec{s} \in \{0, 1\}^N $.  The idea behind this assumption is that because taking the complement of a strand essentially amounts to a relabelling of the bases and a change in the order in which those bases are read, without any kind of specific sequence information there is no reason {\it a priori} to assume that a complementarity symmetry does not hold.  Note that for a strand pair of the form $ (\sigma, \bar{\sigma}) $, we have that $ \gamma (\sigma, \bar{\sigma}) = \tau (\sigma, \bar{\sigma}) $.  Therefore, for genomes consisting of entirely of chromosomes comprised of perfectly complementary strands, we have that $ \gamma^{\vec{s}} \vec{\sigma} = \tau^{\vec{s}} \vec{\sigma} $, and so the complementarity symmetry automatically holds.

We further assume that the transition probability $ p((\sigma_i''; \vec{\sigma}''), (\sigma_i, \sigma_i')) $ obeys a complementarity symmetry, that is, $ p((\gamma^{s_i} \sigma_i''; \gamma^{\vec{s}} \vec{\sigma}''), \gamma^{s_i} (\sigma_i, \sigma_i')) = p((\sigma_i''; \vec{\sigma}''), (\sigma_i, \sigma_i')) $.  Such a 
condition can be accomplished if we assume that mutations are due to a base-independent mismatch probability $ \epsilon_{\vec{\sigma}} $, which obeys the complementarity symmetry.

It may be shown that a population distribution that initially obeys the complementarity symmetry will obey this symmetry for all time, assuming that the fitness landscapes and transition probabilities obey this symmetry.  Because this derivation was already done in \cite{TannHaploid}, we will not repeat it here.  Furthermore, if the population distribution, along with the fitness landscape and transition probabilities, all obey a complementarity symmetry, then we may express the quasispecies equations in a more convenient form.  Again, the derivation has been previously worked out in \cite{TannHaploid}, so we simply present the final results here.  For random segregation, we have,
\begin{eqnarray}
\frac{d y_{\vec{\sigma}}}{dt} 
& = &
-(\bar{\kappa}(t) + \kappa_{\vec{\sigma}}) y_{\vec{\sigma}}
+ \frac{1}{2^{N-1}} 
\sum_{\vec{\sigma}''} \kappa_{\vec{\sigma}''} y_{\vec{\sigma}''}
\times
\nonumber \\
&   & 
\sum_{\vec{s} \in \{0, 1\}^N} 
\prod_{i = 1}^{N} p((\sigma_i''; \vec{\sigma}''), (\gamma \tau)^{s_i} (\sigma_i, \sigma_i'))
\nonumber \\
\end{eqnarray}

For immortal DNA strand co-segregation, we have,
\begin{eqnarray}
\frac{d y_{\vec{\sigma}}}{dt}
& = &
-(\bar{\kappa}(t) + \kappa_{\vec{\sigma}}) y_{\vec{\sigma}}
+ \sum_{\vec{\sigma}''} \kappa_{\vec{\sigma}''} y_{\vec{\sigma}''}
\times 
\nonumber \\
&   &
[\prod_{i = 1}^{N} p((\sigma_i''; \vec{\sigma}''), (\sigma_i, \sigma_i')) +
 \prod_{i = 1}^{N} p((\sigma_i'''; \vec{\sigma}''), (\bar{\sigma}_i, \bar{\sigma}_i'))] 
\nonumber \\
\end{eqnarray}

\subsection{The Infinite Sequence Length Equations}

We now proceed to determine how the random segregation and immortal strand co-segregation equations look in the limit of infinite sequence length.  In doing so, we will consider fitness landscapes that have certain properties that will allow for a considerably simplified version of the equations.  The assumption of infinite sequence length is a common one in quasispecies theory \cite{TannQuasReview}, and is simply a mathematical formalization of the assumption of very long genome lengths.

\subsubsection{The master genome and homologous groups}

To begin, we assume that there exists a ``master" genome, $ \hat{\sigma}_0 = \{\{\sigma_{0, 1}, \bar{\sigma}_{0,1}\}, \dots, \{\sigma_{0, N}, \bar{\sigma}_{0, N}\}\} $, that has the wild-type fitness $ k > 1 $.  This master genome consists of $ M $ distinct strand-pairs, denoted $ \{\sigma_{0, i_1}, \bar{\sigma}_{0, i_1}\}, \dots, \{\sigma_{0, i_M}, \bar{\sigma}_{0, i_M}\} $, where the master genome consists of $ n_k $ pairs of the $ k^{\mbox{th}} $ strand-pair, so that $ N = n_1 + \dots + n_M $.  We define the {\it $ k^{\mbox{th}} $ homologous group} of the master genome to be precisely the $ n_k $ copies of the $ k^{\mbox{th}} $ strand-pair, 
$ \{\sigma_{0, i_k}, \bar{\sigma}_{0, i_k}\} $.

We also let $ L_k $ denote the length, or the number of base-pairs, in $ \{\sigma_{0, i_k}, \bar{\sigma}_{0, i_k}\} $.  The total length, $ L $, of the master genome, is then defined to be $ L = n_1 L_1 + \dots + n_M L_M $.  We then define $ \alpha_k = L_k/L $.

We assume that, during replication, daughter strand synthesis is not error-free, and is characterized by a per-base mismatch probability of $ \epsilon $.  We then allow the total sequence length, $ L $, of the master genome to become infinite, while keeping $ \mu \equiv \epsilon L $ to be constant.  Physically, this corresponds to maintaining a constant replication fidelity in the limit of very large genomes.  This is a common assumption in quasispecies models, and reflects the fact that the average number of mutations per genome per replication cycle, as measured by $ \mu $, is generally far smaller than the size of the genomes themselves \cite{TannQuasReview, WilkeQuasReview, BullQuasReview}.

In the limit of infinite sequence length, we may make the following assumptions about the master genome:  For any two indices $ k \neq l $, we have that,
\begin{eqnarray}
&   &
D_H(\sigma_{0, i_k}, \bar{\sigma}_{0, i_l}) = \infty
\nonumber \\
&   &
D_H(\sigma_{0, i_k}, \bar{\sigma}_{0, i_l}) = \infty, \mbox{ $ k \neq l $}
\end{eqnarray}
where $ D_H(\sigma_1, \sigma_2) $ denotes the {\it Hamming Distance} between any two sequences (the Hamming distance is the number of positions where the two sequences differ).

To understand the basis for these assumptions, we may note that, in the limit of infinite sequence length, a given sequence will, on average, differ from its complement at an infinite number of positions \cite{TannSemiConserve}.  Also, since we do not assume any kind of correlation between the homologous groups, we assume that the strands from distinct homologous groups also differ from each other at an infinite number of positions.

We now consider an initially clonal population consisting entirely of the master genome that is allowed to reproduce and evolve.  After some time, consider some strand-pair $ \{\sigma, \sigma'\} $ in some genome of some organism.  Suppose that this strand pair has the property that, for some $ k $, both Hamming distances $ D_H(\sigma, \sigma_{0, i_k}) $ and $ D_H(\sigma', \bar{\sigma}_{0, i_k}) $ are finite.  In this case, we say that $ \{\sigma, \sigma'\} $ belongs to the 
{\it $ k^{\mbox{th}} $ homologous group}.  Then it follows that $ D_H(\sigma, \sigma_{0, i_l}) $, $ D_H(\sigma, \bar{\sigma}_{0, i_l}) $, $ D_H(\sigma', \sigma_{0, i_l}) $, and $ D_H(\sigma', \bar{\sigma}_{0, i_l}) $ are infinite for $ l \neq k $.  It also follows that $ D_H(\sigma, \bar{\sigma}_{0, i_k}) $ and $ D_H(\sigma', \sigma_{0, i_k}) $ are infinite, and that $ D_H(\sigma, \sigma') $ is infinite.  As a result, a given strand-pair can only belong to at most one homologous group.

When $ \{\sigma, \sigma'\} $ replicates, both $ \sigma $ and $ \sigma' $ act as templates for the synthesis of the complementary daughter strand.  Because $ \mu $, only a finite number of mismatches will occur in both daughter strand syntheses.  As a result, if $ \sigma $ produces $ \sigma_1 $, with daughter $ \sigma_2 $, then we have that $ D_H(\sigma_1, \sigma_{0, i_k}) $ and $ D_H(\sigma_2, \bar{\sigma}_{0, i_k}) $ are both finite.  A similar result holds for the daughter strand-pair produced by $ \sigma' $.  Note then that the daughters of $ \{\sigma, \sigma'\} $ also belong to the $ k^{\mbox{th}} $ homologous group.

Consider a genome where, for $ k = 1, \dots, M $, there are exactly $ n_k $ strand-pairs belonging to the $ k^{\mbox{th}} $ homologous group.  These $ n_k $ strand-pairs produce, upon replication, $ 2 n_k $ strand-pairs belonging to the $ k^{\mbox{th}} $ homologous group, which then segregate equally into two daughter cells, so that each of the daughter cells have exactly $ n_k $ strand-pairs belonging to the $ k^{\mbox{th}} $ homologous group.  By induction, it follows that, if we begin with a clonal population consisting entirely of the master genome, then for all times all genomes in the population will have, for each $ k = 1, \dots, M $, exactly $ n_k $ strand-pairs belonging to the $ k^{\mbox{th}} $ homologous group.

\subsubsection{Viable chromosomes and the fitness landscape}

A given genome is taken to have the wild-type fitness of $ k $ if each homologous group contains at least one functional, or {\it viable}, chromosome.  Otherwise, the fitness is taken to be $ 1 $.  To completely characterize the fitness landscape, we therefore need to properly define what we mean by a ``viable'' chromosome.  So, consider some strand-pair, $ \{\sigma, \sigma'\} $, that belongs to the $ k^{\mbox{th}} $ homologous group.  Then we either have that $ D_H(\sigma, \sigma_{0, i_k}) $, $ D_H(\sigma', \bar{\sigma}_{0, i_k}) $ are finite, or $ D_H(\sigma', \sigma_{0, i_k}) $, $ D_H(\sigma, \bar{\sigma}_{0, i_k}) $ are finite.  Let us assume that the former case holds, since the two cases are completely equivalent.

Then let $ l_C $ denote the number of base-pairs where $ \sigma $ and $ \sigma' $ are complementary, but where $ \sigma $ and $ \sigma' $ differ from $ \sigma_{0, i_k} $ and $ \bar{\sigma}_{0, i_k} $, respectively.  Let $ l_L $ denote the number of base-pairs where $ \sigma $ differs from $ \sigma_{0, i_k} $, but where $ \sigma' $ is identical to $ \bar{\sigma}_{0, i_k} $.  Let $ l_R $ denote the number of base-pairs where $ \sigma $ is identical to $ \sigma_{0, i_k} $, but where $ \sigma' $ differs from $ \bar{\sigma}_{0, i_k} $.  Finally, let $ l_B $ denote the number of base-pairs where both $ \sigma $ and $ \sigma' $ are non-complementary, and differ from $ \sigma_{0, i_k} $ and $ \bar{\sigma}_{0, i_k} $, respectively.

Then the strand-pair $ \{\sigma, \sigma'\} $ is said to be ``viable'' if and only if $ l_C = l_B = 0 $ and $ l_L + l_R \leq l_k^{*} $, where $ l_k^{*} $ is a function of the homologous group number.  The idea here is that if either $ l_C $ or $ l_B $ are positive, then there are regions of the chromosome where sequence information is lost, rendering the chromosome non-functional.  However, where one strand differs from the master sequence but the other strand does not, sequence information is preserved.  If there are not too many such mismatches, or {\it lesions}, then the cellular enzymatic machinery can recover the master sequence information, rendering the chromosome functional.

This fitness landscape is of course a great oversimplification of actual fitness landscapes.  Nevertheless, it is a useful first approximation with which we can obtain analytical results from our evolutionary dynamics equations.

\subsubsection{Population classes, lesion repair, and the infinite sequence length equations}

The master genome gives rise to $ 2^N N!/(n_1! \times \dots \times n_M!) $ ordered strand-pair vectors, given by 
$ (\tau^{s_1}(\sigma_{0, \pi_N(1)}, \bar{\sigma}_{0, \pi_N(1)}), \dots, \tau^{s_N} (\sigma_{0, \pi_N(N)}, \bar{\sigma}_{0, \pi_N(N)})) $, where $ \vec{s} = (s_1, \dots, s_N) \in \{0, 1\}^N $, and $ \pi_N \in \pi_N(\hat{\sigma}_0) $.  We may use this ordering to group the ordered strand-pair vectors into classes, as follows:  First, we pick an ordering for the set of permutations $ \pi_N(\hat{\sigma}_0) $, and list them in some order $ \pi_{N, 1}, \pi_{N, 2}, \dots $.  Also, given a $ \vec{s} \in \{0, 1\}^N $, we define $ k $ to be the number that $ \vec{s} $ represents in binary notation, so that $ k = s_1 2^{N-1} + s_2 2^{N-2} + \dots + s_N $.  

Given an ordered strand-pair vector $ \vec{\sigma} = ((\sigma_1, \sigma_1'), \dots, (\sigma_N, \sigma_N')) $, we say that $ \vec{\sigma} $ belongs to class
$ (n, k) $ if, for each $ i = 1, \dots, N $, we have that $ D_H(\sigma_i, \sigma_{0, \pi_{N, n}(i)}) $, $ D_H(\sigma_i', \bar{\sigma}_{0, \pi_{N, n}(i)}) $ are finite if
$ s_i = 0 $, or $ D_H(\sigma_i, \bar{\sigma}_{0, \pi_{N, n}(i)} $, $ D_H(\sigma_i', \sigma_{0, \pi_{N, n}(i)}) $ are finite if $ s_i = 1 $, where $ (s_1, \dots, s_N) $
is the binary representation of $ k $ as stated above.

We make the following claim:  If we start with a clonal population consisting entirely of the wild-type (i.e. the master genome), then all genomes produced by the evolutionary dynamics of the population give rise to ordered strand-pair vectors belonging to a unique class.  To prove this, we must show that all genomes produced by the evolutionary dynamics give rise to ordered strand-pair vectors belonging to some class, and then we must show that a given ordered strand-pair vector cannot belong to more than one class.

We have already shown that all genomes produced by the evolutionary dynamics of the population give rise to genomes which have $ n_k $ chromosomes belonging to the $ k^{\mbox{th}} $ homologous group for each $ k = 1, \dots, M $.  Let us then consider some ordered strand-pair vector $ \vec{\sigma} =
((\sigma_1, \sigma_1'), \dots, (\sigma_N, \sigma_N')) $ generated by some genome in the population.  The ordered strand-pair $ (\sigma_i, \sigma_i') $ is generated from the strand pair $ \{\sigma_i, \sigma_i'\} $, which in turn belongs to some homologous group as defined above.  We say that $ (\sigma_i, \sigma_i') $ belongs to the same homologous group as $ \{\sigma_i, \sigma_i'\} $.  

So, for each homologous group $ k $, let $ i_{k, 1}, \dots, i_{k, n_k} $ denote the indices of the ordered strand-pairs belonging to the $ k^{\mbox{th}} $ homologous group.  Consider then some $ i \in \{i_{k, 1}, \dots, i_{k, n_k}\} $, and let us consider the pair of Hamming distances $ D_H(\sigma_i, \sigma_{0, \pi_N(i)}) $, $ D_H(\sigma_i', \bar{\sigma}_{0, \pi_N(i)}) $, and $ D_H(\sigma_i, \bar{\sigma}_{0, \pi_N(i)}) $, $ D_H(\sigma_i', \sigma_{0, \pi_N(i)}) $.  If the first pair of Hamming distances are finite, then the second pair is infinite, and vice versa.  However, unless $ \{\sigma_{0, \pi_N(i)}, \bar{\sigma}_{0, \pi_N(i)}\} $ is equal to $ \{\sigma_{0, i_k}, \bar{\sigma}_{0, i_k}\} $, the master ordered strand-pair of the $ k^{\mbox{th}} $ homologous group, then both pairs of Hamming distances are infinite.  Therefore, in order for each ordered strand-pair $ (\sigma_i, \sigma_i') $, where $ i \in \{i_{k, 1}, \dots, i_{k, n_k}\} $ to have the property that either $ D_H(\sigma_i, \sigma_{0, \pi_N(i)}) $, $ D_H(\sigma_i', \bar{\sigma}_{0, \pi_N(i)}) $ or $ D_H(\sigma_i, \bar{\sigma}_{0, \pi_N(i)}) $, $ D_H(\sigma_i', \sigma_{0, \pi_N(i)}) $ are finite, it must follow that $ \pi_N $ must be a permutation that sends the $ n_k $ master strand-pairs associated with the $ k^{\mbox{th}} $ homologous group to the indices $ \{i_{k, 1}, \dots, i_{k, n_k}\} $.  In order for this to hold for all the homologous groups, it follows that $ \pi_N $ must be the unique permutation that sends, for each homologous group $ k $, the $ n_k $ master strand-pairs to the indices $ \{i_{k, 1}, \dots, i_{k, n_k}\} $.  We let $ \pi_{N, n} $ denote this particular permutation, where $ n $ represents the position of this permutation in the ordering of the permutations of $ \pi_N(\hat{\sigma}_0) $.

Now, for a given $ i \in \{i_{k, 1}, \dots, i_{k, n_k}\} $, we have shown that the pair of Hamming distances $ D_H(\sigma_i, \sigma_{0, i_k}) $, $ D_H(\sigma_i', \bar{\sigma}_{0, i_k}) $, and $ D_H(\sigma_i, \bar{\sigma}_{0, i_k}) $, $ D_H(\sigma_i', \sigma_{0, i_k}) $ cannot be simultaneously finite.  If the first pair of Hamming distances is finite, then we have $ s_i = 0 $, while if the second pair is finite then we have $ s_i = 1 $.  If we let $ k $ denote the number that $ (s_1, \dots, s_N) $ represents in binary notation, then we have that $ \vec{\sigma} $ belongs to the class $ (n, k) $.  Note by construction that $ (n, k) $ must be unique.

Let us now consider the random chromosome segregation equations, and let us consider some vector of ordered strand-pairs $ \vec{\sigma} $ belonging to class $ (n, k) $.  If we look at the sum in the equations, we notice that we have a product of terms, each of which is either $ p((\sigma_i''; \vec{\sigma}''), (\sigma_i, \sigma_i')) $ or $ p((\sigma_i''; \vec{\sigma}''), (\bar{\sigma}_i', \bar{\sigma}_i)) $.  For the first probability to be non-zero, we must have that
$ D_H(\sigma_i'', \sigma_i) $ be finite.  This implies that $ \sigma_i'' $ must be a finite Hamming distance away from the same master strand to which $ \sigma_i $ is a finite Hamming distance away, and so the ordered strand-pair with which $ \sigma_i'' $ is associated must belong to the same homologous group as $ (\sigma_i, \sigma_i') $.  If we let $ (n', k') $ denote the class to which $ \vec{\sigma}'' $ belongs, and if we let $ (n, k) $ denote the class to which $ \vec{\sigma} $ belongs, then we must have that $ n' = n $ and $ k' = k $, and so $ \vec{\sigma}'' $ belongs to the same class as $ \vec{\sigma} $.

Now, for the probability $ p((\sigma_i'', \vec{\sigma}''), (\bar{\sigma}_i', \bar{\sigma}_i)) $ to be non-zero, we must have that $ D_H(\sigma_i'', \bar{\sigma}_i') $ is finite.  Since $ \sigma_i' $ is a finite Hamming distance away from the complement of the master strand to which $ \sigma_i $ is a finite Hamming distance away, we have that $ \bar{\sigma}_i' $ is a finite Hamming distance away from the master strand to which $ \sigma_i $ is a finite Hamming distance away, and so $ \sigma_i'' $ is also a finite Hamming distance away from the master strand to which $ \sigma_i $ is a finite Hamming distance away.  Following a similar argument as before, this implies that $ \vec{\sigma}'' $ belongs to the same class of $ \vec{\sigma} $.

As a result, for random chromosome segregation, we need only consider contributions from ordered strand-pair vectors that are in the same class as the daughter ordered strand-pair vector.

Now let us consider immortal strand co-segregation.  For the probability $ p((\sigma_i''; \vec{\sigma}''), (\sigma_i, \sigma_i')) $ to be non-zero, we have that $ D_H(\sigma_i'', \sigma_i) $ must be finite, and so, following a similar argument as before, we obtain that $ \vec{\sigma}'' $ must belong to the same class as $ \vec{\sigma} $.  For the probability $ p((\sigma_i'''; \vec{\sigma}''), (\bar{\sigma}_i, \bar{\sigma}_i')) $ to be non-zero, we must have that $ D_H(\sigma_i''', \bar{\sigma}_i) $ is finite, and so $ \sigma_i''' $ must be a finite Hamming distance away from the complement of the master strand to which $ \sigma_i $ is a finite Hamming distance away.  Therefore, $ \sigma_i'' $ must be a finite Hamming distance away from the master strand to which $ \sigma_i $ is a finite Hamming distance away, and so we obtain that $ \vec{\sigma}'' $ must belong to the same class as $ \vec{\sigma} $.

As a result, for immortal strand co-segregation, we need only consider contributions from ordered strand-pair vectors that are in the same class as the daughter ordered strand-pair vector.

At this point, the random and immortal strand segregation equations for arbitrary genomes become formally identical to the equations for haploid genomes.  Since these equations have already been derived in \cite{TannHaploid}, we obtain, that the infinite sequence length equations are, for random chromosome segregation,
\begin{widetext}
\begin{eqnarray}
&   &
\frac{d z_{((l_{C, 1}, 0, l_1, 0), \dots, (l_{C, N}, 0, l_N, 0))}}{dt} 
= -(\kappa_{((l_{C, 1}, 0, l_1, 0), \dots, (l_{C, N}, 0, l_N, 0))} + \bar{\kappa}(t)) z_{((l_{C, 1}, 0, l_1, 0), \dots, (l_{C, N}, 0, l_N, 0))}
\nonumber \\
&   &
+ \frac{1}{2^{n - 1}} \frac{1}{l_1! \cdots l_N!} e^{-\mu (1 - \lambda/2)} \prod_{i = 1}^{N} [\tilde{\alpha}_i \mu (1 - \lambda)]^{l_i}
\sum_{l_{1, 1}' = 0}^{l_{C, 1}} \frac{1}{l_{1, 1}'!} (\frac{\lambda \tilde{\alpha}_i \mu}{2})^{l_{1, 1}'} \dots \sum_{l_{1, N}' = 0}^{l_{C, N}} \frac{1}{l_{1, N}'!} (\frac{\lambda \tilde{\alpha}_N \mu}{2})^{l_{1, N}'}
\nonumber \\
&   &
\times 
\sum_{l_{2, 1}' = 0}^{l_{C, 1} - l_{1, 1}'} \dots 
\sum_{l_{2, N}' = 0}^{l_{C, N} - l_{1, N}'} \dots
\sum_{l_{3, 1}' = 0}^{\infty} \dots \sum_{l_{3, N}' = 0}^{\infty}
\kappa_{((l_{C, 1} - l_{1, 1}' - l_{2, 1}', l_{2, 1}', l_{3, 1}', 0), \dots, (l_{C, N} - l_{1, N}' - l_{2, N}', l_{2, N}', l_{3, N}', 0))}
\nonumber \\
&   &
\times
z_{((l_{C, 1} - l_{1, 1}' - l_{2, 1}', l_{2, 1}', l_{3, 1}', 0), \dots, (l_{C, N} - l_{1, N}' - l_{2, N}', l_{2, N}', l_{3, N}', 0))}
\end{eqnarray}
\end{widetext}
where $ n $ is the number of strand-pairs with $ l_i > 0 $.  Note that we do not use the $ \alpha_i $ symbol, but rather $ \tilde{\alpha}_i $.  The reason for this is that $ \alpha_i $ refers to $ L_i/L $, where $ L_i $ is the length of the master chromosome of the $ i^{\mbox{th}} $ homologous group.  Here, $ \tilde{\alpha}_i $ refers to $ L_i/L $, where in this case $ L_i $ is the length of the $ i^{\mbox{th}} $ chromosome in the chromosome ordering associated with the given class of vectors of ordered strand-pairs.  If the $ i^{\mbox{th}} $ chromosome belongs to the $ k^{\mbox{th}} $ homologous group, then $ \tilde{\alpha}_i = \alpha_k $.

For immortal strand co-segregation, we have that,
\begin{widetext}
\begin{eqnarray}
&   &
\frac{z_{((l_{C, 1}, 0, l_1, 0), \dots, (l_{C, N}, 0, l_N, 0))}}{dt}
= -(\kappa_{((l_{C, 1}, 0, l_1, 0), \dots, (l_{C, N}, 0, l_N, 0))} + \bar{\kappa}(t)) z_{((l_{C, 1}, 0, l_1, 0), \dots, (l_{C, N}, 0, l_N, 0))}
\nonumber \\
&   &
+ \frac{1}{l_1! \cdots l_N!} e^{-\mu (1 - \lambda/2)} \prod_{i = 1}^{N} [\tilde{\alpha}_i \mu (1 - \lambda)]^{l_i}
\sum_{l_{1, 1}' = 0}^{l_{C, 1}} \frac{1}{l_{1, 1}'!} (\frac{\lambda \tilde{\alpha}_1 \mu}{2})^{l_{1, 1}'} \dots \sum_{l_{1, N}' = 0}^{l_{C, N}} \frac{1}{l_{1, N}'!} 
(\frac{\lambda \tilde{\alpha}_N \mu}{2})^{l_{1, N}'}
\nonumber \\
&   &
\times
\sum_{l_{2, 1}' = 0}^{\infty} \dots \sum_{l_{2, N}' = 0}^{\infty} \kappa_{((l_{C, 1} - l_{1, 1}', 0, l_{2, 1}', 0), \dots, (l_{C, N} - l_{1, N}', 0, l_{2, N}', 0))}
z_{((l_{C, 1} - l_{1, 1}', 0, l_{2, 1}', 0), \dots, (l_{C, N} - l_{1, N}', 0, l_{2, N}', 0))}
\nonumber \\
&   &
+ \frac{1}{l_1! \dots l_N!} e^{-\mu (1 - \lambda/2)} \prod_{i = 1}^{N} [\tilde{\alpha}_i \mu (1 - \lambda)]^{l_i}
\sum_{l_{1, 1}' = 0}^{l_{C, 1}} \frac{1}{l_{1, 1}'!} (\frac{\lambda \tilde{\alpha}_1 \mu}{2})^{l_{1, 1}'} \dots \sum_{l_{1, N}' = 0}^{l_{C, N}} \frac{1}{l_{1, N}'!}
(\frac{\lambda \tilde{\alpha}_N \mu}{2})^{l_{1, N}'}
\nonumber \\
&   &
\times
\sum_{l_{2, 1}' = 0}^{l_{C, 1} - l_{1, 1}'} \dots \sum_{l_{2, N}' = 0}^{l_{C, N} - l_{1, N}'}
\kappa_{((l_{C, 1} - l_{1, 1}' - l_{2, 1}', 0, l_{2, 1}', 0), \dots, (l_{C, N} - l_{1, N}' - l_{2, N}', 0, l_{2, N}', 0))} 
z_{((l_{C, 1} - l_{1, 1}' - l_{2, 1}', 0, l_{2, 1}', 0), \dots, (l_{C, N} - l_{1, N}' - l_{2, N}', 0, l_{2, N}', 0))}
\nonumber \\
\end{eqnarray}
\end{widetext}

Here, we define $ z_{((l_{C, 1}, l_{L, 1}, l_{R, 1}, l_{B, 1}), \dots, (l_{C, N}, l_{L, N}, l_{R, N}, l_{B, N}))} $ to be the fraction of vectors of ordered strand-pairs in the population, belonging to a specific class, characterized by the parameters $ ((l_{C, 1}, l_{L, 1}, l_{R, 1}, l_{B, 1}), \dots, (l_{C, N}, l_{L, N}, l_{R, N}, l_{B, N})) $, where $ l_{C, i}, l_{L, i}, l_{R, i}, l_{B, i}) $ refer to the values of $ l_C, l_L, l_R, l_B $ for the $ i^{\mbox{th}} $ strand-pair, respectively.

The parameter $ \lambda $ is a lesion repair probability, and is the probability that a given mismatch that survived all error repair mechanisms associated with the replication process (e.g. proofreading and mismatch repair) will eventually be eliminated by the lesion repair machinery of the cell.  Here, because there is no longer any discrimination between parent and daughter strands, if a given mismatch is eliminated, then there is a $ 50\% $ probability that the original base-pair will be restored, and a $ 50\% $ probability that a mutation will be fixed in the genome.

For random chromosome segregation, we are able to show in \cite{TannHaploid} that vectors of ordered strand-pairs pairs with $ l_{B, i} > 0 $ cannot be produced through replication, hence we may assume that $ l_{B, i} = 0 $.  Furthermore, we can show that $ l_{L, i}, l_{R, i} $ cannot be simultaneously greater than $ 0 $.  We only show the equations allowing for $ l_{R, i} > 0 $, since the equations where we allow $ l_{L, i} > 0 $ are identical.  For those values of $ i $ for which $ l_{L, i} > 0 $ and $ l_{R, i} = 0 $, we have that $ l_i $ represents the value of $ l_{L, i} $.  The equations that follow are then identical to what is written above.

For immortal strand co-segregation, we are able to show in \cite{TannHaploid} that vectors of ordered strand-pairs with $ l_{B, i} > 0 $ cannot be produced through replication, hence we may assume that $ l_{B, i} = 0 $.  Furthermore, we can show that $ l_{L, i} $ must be $ 0 $ as well.

\subsubsection{Perfect lesion repair}

In contrast to the haploid case, solving for the steady-state mean fitness of the general case turns out to be considerably more difficult.  We have therefore decided to solve for the steady-state mean fitness for the specific case where $ \lambda = 1 $.  This assumes perfect lesion repair, so that we are dealing with genomes where each chromosome consists of perfectly complementary DNA strands.  In this case, it may be shown that both the random and immortal strand co-segregation equations reduce to,
\begin{eqnarray}
&    &
\frac{d z_{(l_1, \dots, l_N)}}{dt} = -(\kappa_{(l_1, \dots, l_N)} + \bar{\kappa}(t)) z_{(l_1, \dots, l_N)}
\nonumber \\
&    &
+ 2 e^{-\mu/2} \sum_{l_1' = 0}^{l_1} \frac{1}{l_1'!} (\frac{\tilde{\alpha}_1 \mu}{2})^{l_1'} \cdots \sum_{l_N' = 0}^{l_N} \frac{1}{l_N'!} (\frac{\tilde{\alpha}_N \mu}{2})^{l_N'}
\nonumber \\
&    &
\times 
\kappa_{(l_1 - l_1', \dots, l_N - l_N')} z_{(l_1 - l_1', \dots, l_N - l_N')}
\end{eqnarray}
where we have changed notation so that $ z_{(l_1, \dots, l_N)} $ in the notation of the previous equation refers to $ z_{((l_1, 0, 0, 0), \dots, (l_N, 0, 0, 0))} $ of the random and immortal strand segregation equations given earlier.

\section{Results and Discussion}

In this section, we will obtain the steady-state mean fitness for the fitness landscape defined in the previous subsection.  For populations within a given class, we begin by defining $ z_{\{i_1, \dots, i_k\}} $ to be the total fraction of vectors of ordered strand-pairs where the chromosomes with indices $ i_1, \dots, i_k $ are non-viable, while the remaining chromosomes are viable.  To define this population fraction more formally, we introduce the following notation:  We let $ \hat{e}_1, \hat{e}_2, \dots, \hat{e}_N $ denote the standard orthonormal basis of $ R^{N} $, so that $ \hat{e}_1 = (1, 0, 0, \dots, 0), \hat{e}_2 = (0, 1, 0, \dots, 0), \dots, \hat{e}_N = (0, 0, \dots, 0, 1) $.  We then have that,
\begin{equation}
z_{\{i_1, \dots, i_k\}} = \sum_{l_{i_1} = 1}^{\infty} \dots \sum_{l_{i_k} = 1}^{\infty} z_{l_{i_1} \hat{e}_1 + \dots + l_{i_k} \hat{e}_k}
\end{equation}

From Appendix B, we may then show that,
\begin{eqnarray}
&   &
\frac{d z_{I = \{i_1, \dots, i_k\}}}{dt} = -(\kappa_{I} + \bar{\kappa}(t)) z_{I}
\nonumber \\
&   &
+ 2 e^{-(\sum_{i \in \{1, \dots, N\}/I} \tilde{\alpha}_i) \mu/2}
\times \nonumber \\
&   &
\sum_{J \subseteq I}
[\prod_{i \in J} (1 - e^{-\tilde{\alpha}_i \mu/2})] \kappa_{I/J} z_{I/J}
\end{eqnarray}
where $ \kappa_{I} $ is defined as the fitness of vectors of ordered strand-pairs where the non-viable chromosomes are of indices $ i_1, \dots, i_k $.

Now, the fitness does not depend on the specific indices that are knocked out, but rather the homologous groups to which each set of indices belong.  If we let $ \Gamma_i $ denote the indices corresponding to the homologous group $ i $, then given a set of knocked out indices $ I $, we may define $ G_i \equiv I \bigcap \Gamma_i $ to be the subset of knocked out indices belonging to homologous group $ i $.  We then have that,
\begin{eqnarray}
&   &
\frac{d z_{G_1 \bigcup \dots \bigcup G_M}}{dt} = -(\kappa_{G_1 \bigcup \dots \bigcup G_M} + \bar{\kappa}(t)) z_{G_1 \bigcup \dots \bigcup G_M}
\nonumber \\
&   &
+ 2 e^{-(1 - m_1 \alpha_1 - \dots - m_M \alpha_M) \mu/2}
\times \nonumber \\
&   &
\sum_{G_1' \subseteq G_1} \dots \sum_{G_M' \subseteq G_M}
[\prod_{n = 1}^{M} (1 - e^{-\alpha_n \mu/2})^{m_n}]
\times \nonumber \\
&    &
\kappa_{G_1/G_1' \bigcup \dots \bigcup G_M/G_M'} z_{G_1/G_1' \bigcup \dots \bigcup G_M/G_M'}
\end{eqnarray} 
where $ m_i $ is the number of indices in $ G_i $, so that $ m_i = o(G_i) $.

Now, define $ z(m_1, \dots, m_M) $ to be the total population fraction of genomes with $ m_i $ knocked-out chromosomes from the $ i^{\mbox{th}} $ homologous group.  That is, $ z(m_1, \dots, m_N) \equiv \sum_{G_1 \subseteq \Gamma_1, o(G_1) = m_1} \dots \sum_{G_M \subseteq \Gamma_M, o(G_M) = m_M} z_{G_1 \bigcup \dots \bigcup G_M} $.  We then have, from Appendix B,
\begin{eqnarray}
&   &
\frac{d z(m_1, \dots, m_M)}{dt} = -(\kappa(m_1, \dots, m_M) + \bar{\kappa}(t)) z(m_1, \dots, m_M)
\nonumber \\
&   &
+ 2 e^{-(1 - m_1 \alpha_1 - \dots - m_M \alpha_M) \mu/2}
\times \nonumber \\
&   &
\sum_{m_1' = 0}^{m_1} \dots \sum_{m_M' = 0}^{m_M}
(1 - e^{-\alpha_1 \mu/2})^{m_1'} \times \dots \times (1 - e^{-\alpha_M \mu/2})^{m_M'}
\times \nonumber \\
&   &
{{n_1 - m_1 + m_1'} \choose m_1'} \times \dots \times {{n_M - m_M + m_M'} \choose m_M'} 
\times \nonumber \\
&   &
\kappa(m_1 - m_1', \dots, m_M - m_M') z(m_1 - m_1', \dots, m_M - m_M')
\end{eqnarray}

Now, at steady-state, let $ m^{*} $ denote the smallest value of $ m_1 + \dots + m_M $ for which there exists a $ z(m_1, \dots, m_M) > 0 $.  Then given 
$ m_1, \dots, m_M $ for which $ z(m_1, \dots, m_M) > 0 $ and $ m^{*} = m_1 + \dots + m_M $, we have, at steady-state,
\begin{eqnarray}
&   &
0 = [(2 e^{-(1 - m_1 \alpha_1 - \dots - m_M \alpha_M) \mu/2} - 1) \kappa(m_1, \dots, m_M) - \bar{\kappa}] 
\times \nonumber \\
&   &
z(m_1, \dots, m_M)
\end{eqnarray}
which implies that $ \bar{\kappa} = \kappa(m_1, \dots, m_M) (2 e^{-(1 - m_1 \alpha_1 - \dots - m_M \alpha_M) \mu/2} - 1) $.  The reason for this is that, in the sum in Eq. (24), if $ z(m_1 - m_1', \dots, m_M - m_M') > 0 $, then by definition of $ m^{*} $ we have that $ (m_1 - m_1') + \dots + (m_M - m_M') \geq m^{*} \Rightarrow
m^{*} - (m_1' + \dots + m_M') \geq m^{*} \Rightarrow m_1' + \dots + m_M' \leq 0 \Rightarrow m_1' = \dots = m_M' = 0 $.  

Now, we also have, from Eq. (24), for arbitrary values of $ m_1, \dots, m_M $, that,
\begin{eqnarray}
&   &
\frac{d z(m_1, \dots, m_M)}{dt} \geq [\kappa(m_1, \dots, m_M) 
\times \nonumber \\
&   &
(2 e^{-(1 - m_1 \alpha_1 - \dots - m_M \alpha_M) \mu/2} - 1) - \bar{\kappa}(t)] z(m_1, \dots, m_M)
\nonumber \\
\end{eqnarray}
and so, for the steady-state to be stable, we must have that $ \bar{\kappa} \geq \kappa(m_1, \dots, m_M) (2 e^{-(1 - m_1 \alpha_1 - \dots - m_M \alpha_M) \mu/2} - 1) $.  Combined with the fact that equality holds for some set of values of $ m_1, \dots, m_M $, we then have that,
\begin{eqnarray}
\bar{\kappa} 
& = & 
\max\{\kappa(m_1, \dots, m_M) (2 e^{-(1 - m_1 \alpha_1 - \dots - m_M \alpha_M) \mu/2} - 1)\} 
\nonumber \\
& = &
\max\{k (2 e^{(\alpha_1 + \dots + \alpha_M) \mu/2} - 1), 1\}
\end{eqnarray}

We compared the results of our analysis with results obtained from stochastic simulations of replicating populations.  These are shown in Figures 1-3.  Note the excellent agreement between the analytical expression for the mean fitness and the numerical results obtained from the stochastic simulations.

\begin{figure}
\includegraphics[width = 0.7\linewidth, angle = -90]{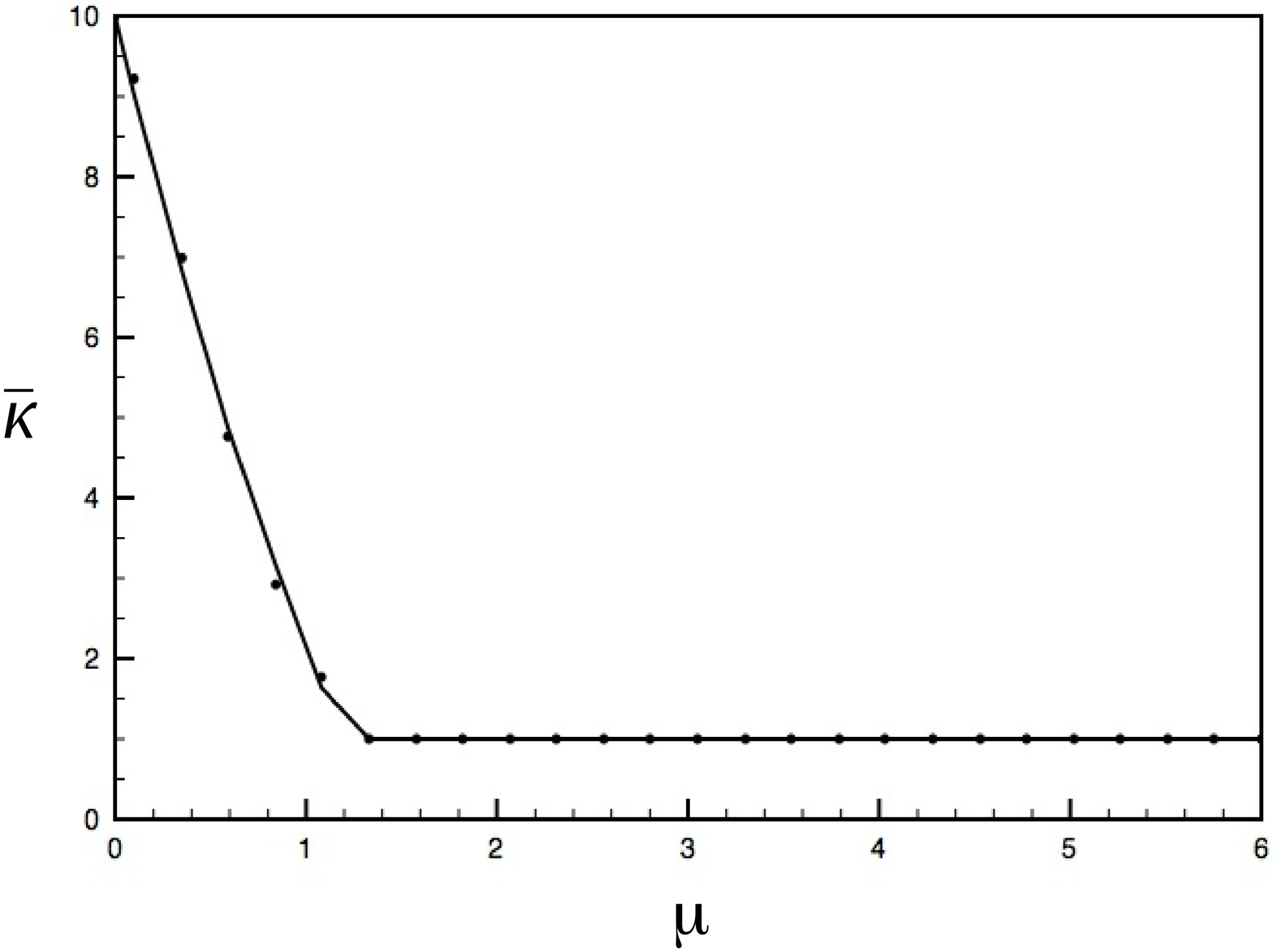}
\caption{A plot of $ \bar{\kappa} $ versus $ \mu $ comparing both the analytical expression for $ \bar{\kappa} $ (solid line) with the values obtained from stochastic simulations (dots).  We have $ M = 2 $, $ n_1 = n_2 = 1 $, $ L_1 = L_2 = 10 $.  We took a population size of $ 1,000 $.}
\end{figure}

\begin{figure}
\includegraphics[width = 0.7\linewidth, angle = -90]{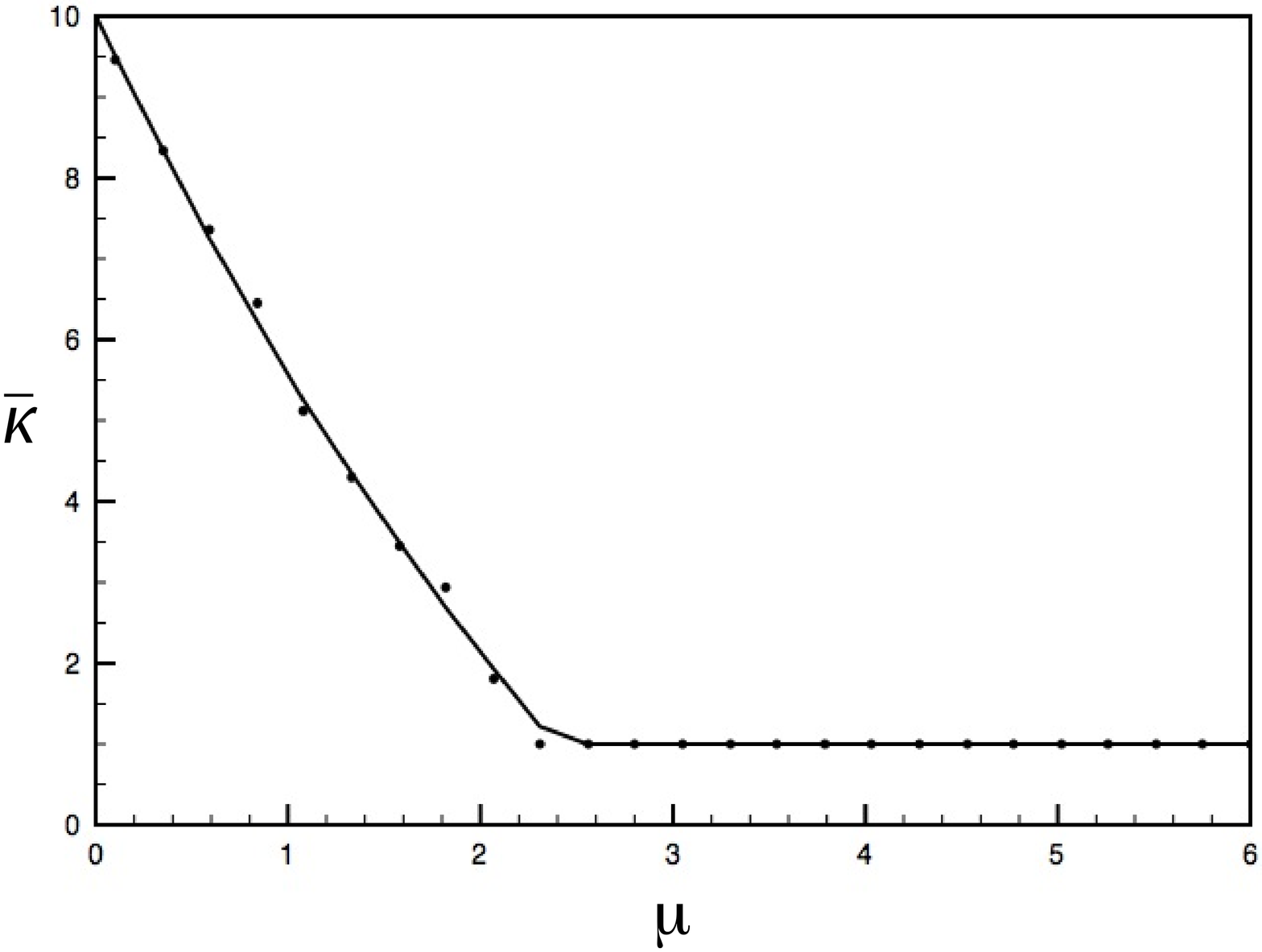}
\caption{A plot of $ \bar{\kappa} $ versus $ \mu $ comparing both the analytical expression for $ \bar{\kappa} $ (solid line) with the values obtained from stochastic simulations (dots).  We have $ M = 2 $, $ n_1 = n_2 = 2 $, $ L_1 = L_2 = 10 $.  We took a population size of $ 1,000 $.}
\end{figure}

\begin{figure}
\includegraphics[width = 0.7\linewidth, angle = -90]{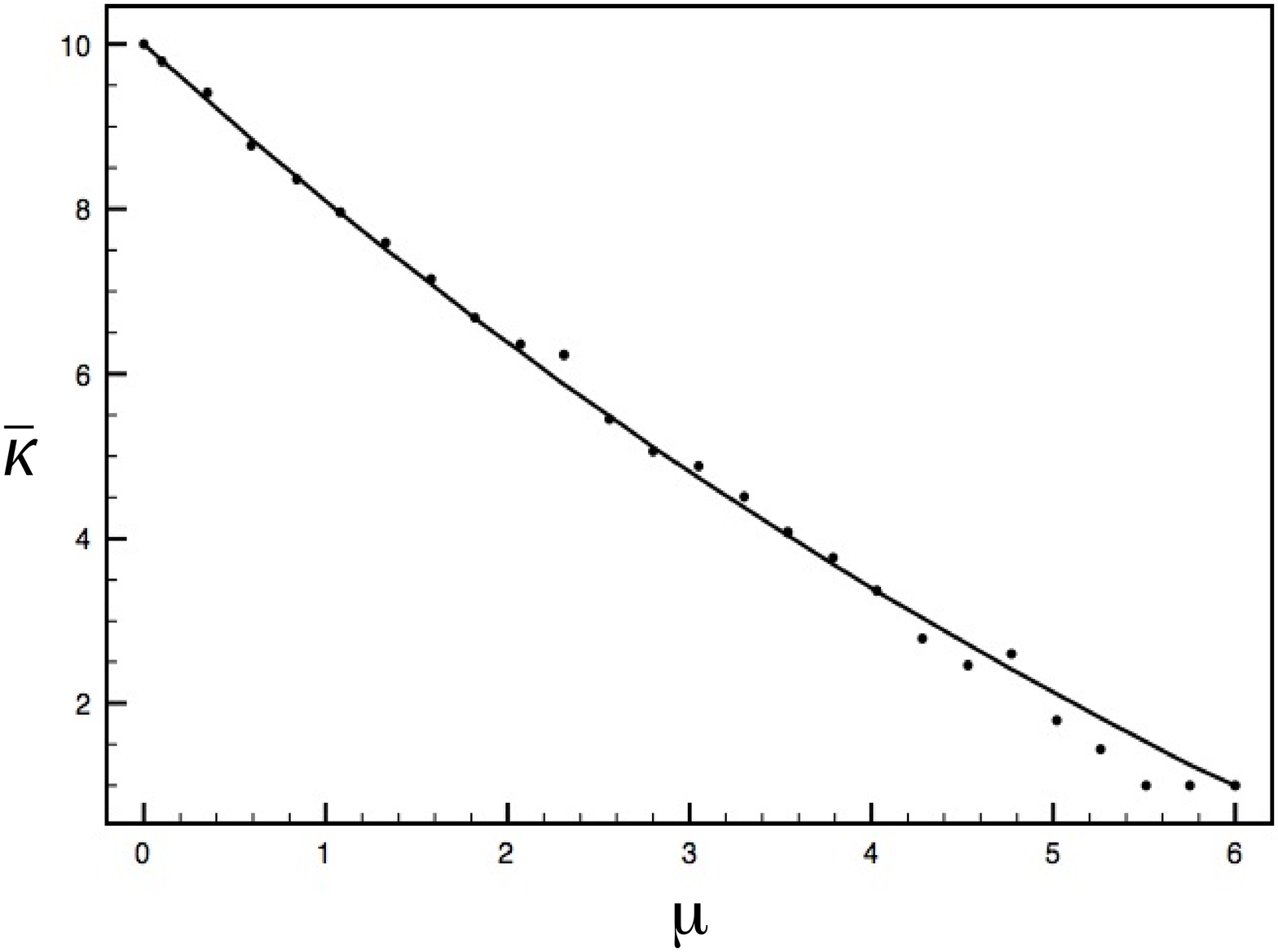}
\caption{A plot of $ \bar{\kappa} $ versus $ \mu $ comparing both the analytical expression for $ \bar{\kappa} $ (solid line) with the values obtained from stochastic simulations (dots).  We have $ M = 2 $, $ n_1 = n_2 = 5 $, $ L_1 = L_2 = 10 $.  We took a population size of $ 1,000 $.}
\end{figure}

\section{Conclusions}

This paper developed the semiconservative quasispecies equations for polysomic genomes.  In contrast to previous work \cite{TannHaploid}, the quasispecies equations developed here are not restricted to haploid genomes, but rather may applied to diploid and even polyploid genomes.  

By an appropriate transformation of variables, these generalized equations may be recast into a form that makes them formally identical to the equations developed for haploid genomes.  However, because of the existence of identical copies of chromosomes in polyploid genomes, we were unable to obtain an analytical expression for the mean fitness for the case of arbitrary lesion repair, as we were able to do for the haploid equations.  This of course does not mean that an analytical expression does not exist.  It simply means that obtaining an analytical expression for the mean fitness is considerably more difficult for the polyploid case than it is for the haploid case.  We therefore solved for the mean fitness for the case of perfect lesion repair ($ \lambda = 1 $), which was considerably more tractable than the general case, leaving the case of arbitrary lesion repair for future work.

The mean fitness results obtained from stochastic simulations were found to be in excellent agreement with the analytical results that we derived for the fitness landscape that was considered in this paper.  We find that beyond a critical mutation rate, the population becomes entirely non-viable, with a low fitness of $ 1 $.  This signals the onset of the error catastrophe, whereby natural selection can no longer localize the population about the fast replicating genotypes, and the result is dynamics governed by pure genetic drift.

In the end, we regard this work as a ``methodology" paper, in the sense that its purpose it to extend the quasispecies formalism developed for haploid genomes to deal with more complicated genomes.  The analytical solution obtained for our chosen fitness landscape and lesion repair probability, along with the stochastic simulation results, are meant to confirm the validity of the master equations (Eqs. (7) and (13)).  Therefore, future research will involve using these equations, along with similar equations developed in \cite{TannHaploid, TannStemCell}, to model the evolutionary dynamics in asexually replicating unicellular populations.  In particular, these equations could be highly useful for modeling mutation-propagation in stem cells and tumors, and could therefore be relevant for cancer modeling and aging.  In this vein, an additional extension of our model that will need to be considered is the incorporation of genomic instability into our framework.

\begin{acknowledgments}

This research was supported by the Israel Science Foundation and by the United States - Israel Binational Science Foundation.  

\end{acknowledgments}

\begin{appendix}

\section{Derivation of the Finite Sequence Length Equations}

In this Appendix, we derive the finite sequence length equations for both random segregation and immortal strand co-segregation.

\subsection{Random segregation}

We begin with random segregation, and our goal is to initially derive equations for the $ x_{\hat{\sigma}} $ population fractions.  Although the chromosomes are formally indistinguishable, for purposes of the derivation we can assign an arbitrary chromosome ordering to every genome.  The only requirement is that once an ordering is chosen, we use that same ordering for the given genome.  Similarly, for each chromosome in a given genome with an assigned ordering, we can tag one of the strands with a ``0'', and the other with a ``1''.  Again, this tagging scheme is arbitrary; however, once chosen, must be consistent.  This chromosome ordering and strand tagging scheme allows us to appropriately keep track of the chromosomes during the replication process.

During the replication process, every strand of every chromosome serves as the template for the synthesis of a daughter strand, and therefore of a new chromosome.  For convenience of the derivation, we assume that each new chromosome segregates into a left daughter cell and a right daughter cell (relevant figures may be found in \cite{TannHaploid}).  For a given parent chromosome from the original cell, the chromosomes formed from the ``0'' strands and the ``1'' strands segregate into opposite cells.  Since chromosome segregation is random, each chromosome has a $ 50\% $ probability of segregating into a given daughter cell.

Note that for a parent genome $ \{\{\sigma_1''^{(0)}, \sigma_1''^{(1)}\}, \dots, \{\sigma_N''^{(0)}, \sigma_N''^{(1)}\}\} $, a given set of parent strands
$ \sigma_1''^{(s_1)}, \dots, \sigma_N''^{(s_N)} $, with each $ s_i = 0,1 $, can only produce the genome $ \{\{\sigma_1, \sigma_1'\}, \dots, \{\sigma_N, \sigma_N'\}\} $ if the parent strands $ \sigma_1''^{(s_1)}, \dots, \sigma_N''^{(s_N)} $ respectively produce $ \{\sigma_{\pi_N(1)}, \sigma_{\pi_N(1)}'\}, \dots, \{\sigma_{\pi_N(N)}, 
\sigma_{\pi_N(N)}'\} $, where $ \pi_N $ denotes a permutation of the strand indices.  Note that when considering the set of permutations of the strand indices, we have to consider those $ \pi_N \in \pi_N(\hat{\sigma}) $, where $ \pi_N(\hat{\sigma}) $ denotes the set of all permutations of the strand indices so that all the ordered strand-pair vectors $ (\{\sigma_{\pi_N(1)}, \sigma_{\pi_N(1)}'\}, \dots, \{\sigma_{\pi_N(N)}, \sigma_{\pi_N(N)}'\}) $ are distinct.  If we consider all permutations, then since some chromosomes are identical, we will be over-counting the total contribution.  We then have,
\begin{widetext}
\begin{eqnarray}
\frac{d x_{\hat{\sigma}}}{dt}
& = &
-(\bar{\kappa}(t) + \kappa_{\hat{\sigma}}) x_{\hat{\sigma}}
+ \sum_{\hat{\sigma}'' = \{\{\sigma_1''^{(0)}, \sigma_1''^{(1)}\}, \dots, \{\sigma_N''^{(0)}, \sigma_N''^{(1)}\}\}}
\kappa_{\hat{\sigma}''} x_{\hat{\sigma}''}
\times
\nonumber \\
&   &
\frac{1}{2^N}
\sum_{s_1 = 0}^{1} \cdots \sum_{s_N = 0}^{1}
\sum_{\pi_N \in \pi_N(\hat{\sigma})}
[\prod_{i = 1}^{N} p((\sigma_i''^{(s_i)}; \hat{\sigma}''), \{\sigma_{\pi_N(i)}, \sigma_{\pi_N(i)}'\}) +
 \prod_{i = 1}^{N} p((\sigma_i''^{(s_i+1)}; \hat{\sigma}''), \{\sigma_{\pi_N(i)}, \sigma_{\pi_N(i)}'\})]
\nonumber \\
& = &
-(\bar{\kappa}(t) + \kappa_{\hat{\sigma}}) x_{\hat{\sigma}}
+ \sum_{\hat{\sigma}'' = \{\{\sigma_1''^{(0)}, \sigma_1''^{(1)}\}, \dots, \{\sigma_N''^{(0)}, \sigma_N''^{(1)}\}\}}
\kappa_{\hat{\sigma}''} x_{\hat{\sigma}''}
\nonumber \\
&   &
\times
\frac{1}{2^{N-1}}
\sum_{\pi_N \in \pi_N(\hat{\sigma})} \sum_{s_1 = 0}^{1} \cdots \sum_{s_N = 0}^{1}
\prod_{i = 1}^{N} p((\sigma_i''^{(s_i)}; \hat{\sigma}''), \{\sigma_{\pi_N(i)}, \sigma_{\pi_N(i)}'\})
\nonumber \\
& = &
-(\bar{\kappa}(t) + \kappa_{\hat{\sigma}}) x_{\hat{\sigma}}
+ \frac{1}{2^{N-1}}
\sum_{\hat{\sigma}'' = \{\{\sigma_1''^{(0)}, \sigma_1''^{(1)}\}, \dots, \{\sigma_N''^{(0)}, \sigma_N''^{(1)}\}\}}
\kappa_{\hat{\sigma}''} x_{\hat{\sigma}''} \times
\nonumber \\
&   &
\sum_{\pi_N \in \pi_N(\hat{\sigma})} 
\prod_{i = 1}^{N}
[p((\sigma_i''^{(0)}; \hat{\sigma}''), \{\sigma_{\pi_N(i)}, \sigma_{\pi_N(i)}'\}) +
 p((\sigma_i''^{(1)}; \hat{\sigma}''), \{\sigma_{\pi_N(i)}, \sigma_{\pi_N(i)}'\})]
\end{eqnarray}
\end{widetext}
which is equivalent to Eq. (3).

We next derive the equations for the ordered chromosome representation.  We obtain,
\begin{widetext}
\begin{eqnarray}
&   &
\frac{d x_{\tilde{\sigma}}}{dt} = -(\bar{\kappa}(t) + \kappa_{\tilde{\sigma}}) x_{\tilde{\sigma}} 
+ \frac{1}{2^{N-1}} \sum_{\hat{\sigma}''} \kappa_{\tilde{\sigma}''} \frac{N!}{n_1'! \times \dots \times n_{m'}'!} x_{\tilde{\sigma}''}
\frac{n_1! \times \dots \times n_m!}{N!}
\times \nonumber \\
&   &
\sum_{\pi_N \in \pi_N(\hat{\sigma})} \prod_{i = 1}^{N}
[p((\sigma_i''; \tilde{\sigma}''), \{\sigma_{\pi_N(i)}, \sigma_{\pi_N(i)}'\}) + p((\sigma_i'''; \tilde{\sigma}''), \{\sigma_{\pi_N(i)}, \sigma_{\pi_N(i)}'\})]
\nonumber \\
&   &
= -(\bar{\kappa}(t) + \kappa_{\tilde{\sigma}}) x_{\tilde{\sigma}}
+ \frac{1}{2^{N-1} N!}
\sum_{\hat{\sigma}''} \sum_{\pi_N' \in \pi_N(\hat{\sigma}'')}
\kappa_{(\{\sigma_{\pi_N'(1)}'', \sigma_{\pi_N'(1)}'''\}, \dots, \{\sigma_{\pi_N'(N)}'', \sigma_{\pi_N'(N)}'''\})}
x_{(\{\sigma_{\pi_N'(1)}'', \sigma_{\pi_N'(1)}'''\}, \dots, \{\sigma_{\pi_N'(N)}'', \sigma_{\pi_N'(N)}'''\})}
\times \nonumber \\
&   &
\sum_{\pi_N \in \pi_N(\hat{\sigma})} n_1! \times \dots \times n_m! \prod_{i = 1}^{N}
[p((\sigma_i''; \tilde{\sigma}''), \{\sigma_{\pi_N(i)}, \sigma_{\pi_N(i)}'\}) + p((\sigma_i'''; \tilde{\sigma}''), \{\sigma_{\pi_N(i)}, \sigma_{\pi_N(i)}'\})]
\nonumber \\
&   &
= -(\bar{\kappa}(t) + \kappa_{\tilde{\sigma}}) x_{\tilde{\sigma}}
+ \frac{1}{2^{N-1} N!}
\sum_{\hat{\sigma}''} \sum_{\pi_N' \in \pi_N(\hat{\sigma}'')}
\kappa_{(\{\sigma_{\pi_N'(1)}'', \sigma_{\pi_N'(1)}'''\}, \dots, \{\sigma_{\pi_N'(N)}'', \sigma_{\pi_N'(N)}'''\})}
x_{(\{\sigma_{\pi_N'(1)}'', \sigma_{\pi_N'(1)}'''\}, \dots, \{\sigma_{\pi_N'(N)}'', \sigma_{\pi_N'(N)}'''\})}
\times \nonumber \\
&   &
\sum_{\pi_N} \prod_{i = 1}^{N} [p((\sigma_{\pi_N'(i)}''; \tilde{\sigma}''), \{\sigma_{\pi_N \pi_N'(i)}, \sigma_{\pi_N \pi_N'(i)}'\}) + p((\sigma_{\pi_N'(i)}'''; \tilde{\sigma}''), \{\sigma_{\pi_N \pi_N'(i)}, \sigma_{\pi_N \pi_N'(i)}'\})]
\nonumber \\
&   &
= -(\bar{\kappa}(t) + \kappa_{\tilde{\sigma}}) x_{\tilde{\sigma}}
+ \frac{1}{2^{N-1}} \frac{1}{N!}
\sum_{\hat{\sigma}''} \sum_{\pi_N' \in \pi_N(\hat{\sigma}'')}
\kappa_{(\{\sigma_{\pi_N'(1)}'', \sigma_{\pi_N'(1)}'''\}, \dots, \{\sigma_{\pi_N'(N)}'', \sigma_{\pi_N'(N)}'''\})}
x_{(\{\sigma_{\pi_N'(1)}'', \sigma_{\pi_N'(1)}'''\}, \dots, \{\sigma_{\pi_N'(N)}'', \sigma_{\pi_N'(N)}'''\})}
\times \nonumber \\
&   &
\sum_{\pi_N} \prod_{i = 1}^{N} [p((\sigma_{\pi_N'(i)}''; \tilde{\sigma}''), \{\sigma_{\pi_N(i)}, \sigma_{\pi_N(i)}'\}) + p((\sigma_{\pi_N'(i)}'''; \tilde{\sigma}''), \{\sigma_{\pi_N(i)}, \sigma_{\pi_N(i)}'\})]
\nonumber \\
&   &
= -(\bar{\kappa}(t) + \kappa_{\tilde{\sigma}}) x_{\tilde{\sigma}}
+ \frac{1}{2^{N-1} N!}
\sum_{\tilde{\sigma}''} \kappa_{\tilde{\sigma}''} x_{\tilde{\sigma}''}
\sum_{\pi_N} \prod_{i = 1}^{N}
[p((\sigma_i''; \tilde{\sigma}''), \{\sigma_{\pi_N(i)}, \sigma_{\pi_N(i)}'\}) + p((\sigma_i'''; \tilde{\sigma}''), \{\sigma_{\pi_N(i)}, \sigma_{\pi_N(i)}'\})]
\nonumber \\
&   &
= -(\bar{\kappa}(t) + \kappa_{\tilde{\sigma}}) x_{\tilde{\sigma}}
+ \frac{1}{2^{N-1} N!}
\sum_{\pi_N} \sum_{\tilde{\sigma}''}
\kappa_{(\{\sigma_{\pi_N^{-1}(1)}'', \sigma_{\pi_N^{-1}(1)}'''\}, \dots, \{\sigma_{\pi_N^{-1}(N)}'', \sigma_{\pi_N^{-1}(N)}'''\})}
x_{(\{\sigma_{\pi_N^{-1}(1)}'', \sigma_{\pi_N^{-1}(1)}'''\}, \dots, \{\sigma_{\pi_N^{-1}(N)}'', \sigma_{\pi_N^{-1}(N)}'''\})}
\times \nonumber \\
&   &
\prod_{i = 1}^{N}
[p((\sigma_{\pi_N^{-1}(i)}''; \tilde{\sigma}''), \{\sigma_i, \sigma_i'\}) + p((\sigma_{\pi_N^{-1}(i)}'''; \tilde{\sigma}''), \{\sigma_i, \sigma_i'\})]
\nonumber \\
&   &
= -(\bar{\kappa}(t) + \kappa_{\tilde{\sigma}}) x_{\tilde{\sigma}} + \frac{1}{2^{N-1}} \sum_{\tilde{\sigma}''} \kappa_{\tilde{\sigma}''} x_{\tilde{\sigma}''}
\prod_{i = 1}^{N}
[p((\sigma_i''; \tilde{\sigma}''), \{\sigma_i, \sigma_i'\}) + p((\sigma_i'''; \tilde{\sigma}''), \{\sigma_i, \sigma_i'\})]
\nonumber \\
\end{eqnarray}
\end{widetext}
where $ \sum_{\pi_N} $ denotes the sum over all permutations of the indices $ 1, \dots, N $.  From these equations, which are equivalent to Eq. (5), the passage to the vector of ordered strand-pairs formulation of the dynamics is identical to the derivation in \cite{TannHaploid}.

\subsection{Immortal strand co-segregation}

For immortal strand segregation, we initially have,
\begin{widetext}
\begin{eqnarray}
&   &
\frac{d x_{\hat{\sigma}^{(N/N)}}}{dt} = -(\kappa_{\hat{\sigma}} + \bar{\kappa}(t)) x_{\hat{\sigma}^{(N/N)}}
\nonumber \\
&   &
\frac{d x_{\hat{\sigma}^{(T/N)}}}{dt} = -(\kappa_{\hat{\sigma}} + \bar{\kappa}(t)) x_{\hat{\sigma}^{(T/N)}}
+ \sum_{\hat{\sigma}''^{(N/N)}} \kappa_{\hat{\sigma}''} x_{\hat{\sigma}''^{(N/N)}} \frac{1}{2^N}
\times \nonumber \\
&   &
\sum_{s_1 = 0}^{1} \dots \sum_{s_N = 0}^{1}
\sum_{\pi_N \in \pi_N(\hat{\sigma}^{(T/N)})} 
[\prod_{i = 1}^{N} p((\sigma_i''^{(s_i)}; \hat{\sigma}''), (\sigma_{\pi_N(i)}, \sigma_{\pi_N(i)}')) + \prod_{i = 1}^{N} p((\sigma_i''^{(s_i + 1)}; \hat{\sigma}''), (\sigma_{\pi_N(i)}, \sigma_{\pi_N(i)}'))]
\nonumber \\
&   &
+ \sum_{\hat{\sigma}''^{(T/N)}} \kappa_{\hat{\sigma}''} x_{\hat{\sigma}''^{(T/N)}}
\sum_{\pi_N \in \pi_N(\hat{\sigma}^{(T/N)})}
[\prod_{i = 1}^{N} p((\sigma_i''^{(0)}; \hat{\sigma}''), (\sigma_{\pi_N(i)}, \sigma_{\pi_N(i)}')) + \prod_{i = 1}^{N} p((\sigma_i''^{(1)}; \hat{\sigma}''), (\sigma_{\pi_N(i)}, \sigma_{\pi_N(i)}'))]
\nonumber \\
&   &
=  -(\kappa_{\hat{\sigma}} + \bar{\kappa}(t)) x_{\hat{\sigma}^{(T/N)}} + \frac{1}{2^{N - 1}} \sum_{\hat{\sigma}''^{(N/N)}} \kappa_{\hat{\sigma}''} x_{\hat{\sigma}''^{(N/N)}}
\times \nonumber \\
&   &
\sum_{\pi_N \in \pi_N(\hat{\sigma}^{(T/N)})} 
[\prod_{i = 1}^{N} [p((\sigma_i''^{(0)}; \hat{\sigma}''), (\sigma_{\pi_N(i)}, \sigma_{\pi_N(i)}')) + p((\sigma_i''^{(1)}; \hat{\sigma}''), (\sigma_{\pi_N(i)}, \sigma_{\pi_N(i)}'))]
\nonumber \\
&   &
+ \sum_{\hat{\sigma}''^{(T/N)}} \kappa_{\hat{\sigma}''} x_{\hat{\sigma}''^{(T/N)}}
\sum_{\pi_N \in \pi_N(\hat{\sigma}^{(T/N)})}
[\prod_{i = 1}^{N} p((\sigma_i''^{(0)}; \hat{\sigma}''), (\sigma_{\pi_N(i)}, \sigma_{\pi_N(i)}')) + \prod_{i = 1}^{N} p((\sigma_i''^{(1)}; \hat{\sigma}''), (\sigma_{\pi_N(i)}, \sigma_{\pi_N(i)}'))]
\end{eqnarray}
\end{widetext}
which is equivalent to Eq. (8).

Converting to the ordered strand-pair formulation of the dynamics, we have,
\begin{widetext}
\begin{eqnarray}
&   &
\frac{d x_{\tilde{\sigma}^{(N/N)}}}{dt} = -(\kappa_{\tilde{\sigma}} + \bar{\kappa}(t)) x_{\tilde{\sigma}^{(N/N)}}
\nonumber \\
&   &
\frac{d x_{\tilde{\sigma}^{(T/N)}}}{dt} =  -(\kappa_{\tilde{\sigma}} + \bar{\kappa}(t)) x_{\tilde{\sigma}^{(T/N)}}
+ \frac{1}{2^{N - 1}} \sum_{\tilde{\sigma}''^{(N/N)}} \kappa_{\tilde{\sigma}''} x_{\tilde{\sigma}''^{(N/N)}} 
\prod_{i = 1}^{N} [p((\sigma_i''; \tilde{\sigma}''), (\sigma_i, \sigma_i')) + p((\sigma_i'''; \tilde{\sigma}''), (\sigma_i, \sigma_i'))]
\nonumber \\
&   &
+ \frac{1}{N!}
\sum_{\hat{\sigma}''^{(T/N)}}
\kappa_{\tilde{\sigma}''} \frac{N!}{n_1'! \times \dots \times n_{m'}'!} x_{\tilde{\sigma}''^{(T/N)}}
\sum_{\pi_N \in \pi_N(\hat{\sigma}^{(T/N)})} n_1! \times \dots \times n_m!
\times \nonumber \\
&   &
[\prod_{i = 1}^{N} p((\sigma_i''^{(0)}; \hat{\sigma}''^{(T/N)}), (\sigma_{\pi_N(i)}, \sigma_{\pi_N(i)}')) + \prod_{i = 1}^{N} p((\sigma_i''^{(1)}; \hat{\sigma}''^{(T/N)}), (\sigma_{\pi_N(i)}, \sigma_{\pi_N(i)}'))]
\nonumber \\
&   &
= -(\kappa_{\tilde{\sigma}} + \bar{\kappa}(t)) x_{\tilde{\sigma}^{(T/N)}}
+ \frac{1}{2^{N - 1}} \sum_{\tilde{\sigma}''^{(N/N)}} \kappa_{\tilde{\sigma}''} x_{\tilde{\sigma}''^{(N/N)}} 
\prod_{i = 1}^{N} [p((\sigma_i''; \tilde{\sigma}''), (\sigma_i, \sigma_i')) + p((\sigma_i'''; \tilde{\sigma}''), (\sigma_i, \sigma_i'))]
\nonumber \\
&   &
+ \frac{1}{N!}
\sum_{\hat{\sigma}''^{(T/N)}} \sum_{\pi_N' \in \pi_N(\hat{\sigma}''^{(T/N)})}
\kappa_{(\{\sigma_{\pi_N'(1)}''^{(0)}, \sigma_{\pi_N'(1)}''^{(1)}\}, \dots, \{\sigma_{\pi_N'(N)}''^{(0)}, \sigma_{\pi_N'(N)}''^{(1)}\})} 
x_{(\{\sigma_{\pi_N'(1)}''^{(0)}, \sigma_{\pi_N'(1)}''^{(1)}\}, \dots, \{\sigma_{\pi_N'(N)}''^{(0)}, \sigma_{\pi_N'(N)}''^{(1)}\})}
\times \nonumber \\
&   &
\sum_{\pi_N} 
[\prod_{i = 1}^{N} p((\sigma_i''^{(0)}; \hat{\sigma}''), (\sigma_{\pi_N(i)}, \sigma_{\pi_N(i)}')) 
+ 
 \prod_{i = 1}^{N} p((\sigma_i''^{(1)}; \hat{\sigma}''), (\sigma_{\pi_N(i)}, \sigma_{\pi_N(i)}'))]
\nonumber \\
&   &
= -(\kappa_{\tilde{\sigma}} + \bar{\kappa}(t)) x_{\tilde{\sigma}^{(T/N)}}
+ \frac{1}{2^{N - 1}} \sum_{\tilde{\sigma}''^{(N/N)}} \kappa_{\tilde{\sigma}''} x_{\tilde{\sigma}''^{(N/N)}} 
\prod_{i = 1}^{N} [p((\sigma_i''; \tilde{\sigma}''), (\sigma_i, \sigma_i')) + p((\sigma_i'''; \tilde{\sigma}''), (\sigma_i, \sigma_i'))]
\nonumber \\
&   &
+ 
\sum_{\tilde{\sigma}''^{(T/N)}} \kappa_{\tilde{\sigma}''} x_{\tilde{\sigma}''^{(T/N)}}
[\prod_{i = 1}^{N} p((\sigma_i''^{(0)}; \tilde{\sigma}''), (\sigma_i, \sigma_i')) 
+ 
 \prod_{i = 1}^{N} p((\sigma_i''^{(1)}; \tilde{\sigma}''), (\sigma_i, \sigma_i'))]
\end{eqnarray}
\end{widetext}
which is equivalent to Eq. (9).  Following the derivation in \cite{TannHaploid}, we may then obtain Eq. (13).

\section{Derivation Details for the Steady-State Mean Fitness}

Using the definition for $ z_{\{i_1, \dots, i_k\}} $ provided in the main text, and from the infinite sequence length equations, we have that,
\begin{eqnarray}
&    &
\frac{d z_{\{i_1, \dots, i_k\}}}{dt} = -(\kappa_{\{i_1, \dots, i_k\}} + \bar{\kappa}(t)) z_{\{i_1, \dots, i_k\}} 
+ 2 e^{-\mu/2} 
\times \nonumber \\
&   &
\sum_{l_{i_1} = 1}^{\infty} \dots \sum_{l_{i_k} = 1}^{\infty}
\sum_{l_{i_1}' = 0}^{l_{i_1}} \dots \sum_{l_{i_k}' = 0}^{l_{i_k}}
\prod_{i \in \{i_1, \dots, i_k\}} \frac{1}{l_i'!} (\frac{\lambda \alpha_i \mu}{2})^{l_i'}
\times \nonumber \\
&   &
\kappa_{(l_{i_1} - l_{i_1}') \hat{e}_{i_1} + \dots + (l_{i_k} - l_{i_k}') \hat{e}_{i_k}} 
z_{(l_{i_1} - l_{i_1}') \hat{e}_{i_1} + \dots + (l_{i_k} - l_{i_k}') \hat{e}_{i_k}} 
\nonumber \\
\end{eqnarray}

We now make use of the following identity:
\begin{eqnarray}
&    &
\sum_{l_1' = 0}^{l_1} \dots \sum_{l_n' = 0}^{l_n} f(l_1', \dots, l_n') = \sum_{\{j_1, \dots, j_k\} \subseteq \{1, \dots, n\}}
\times \nonumber \\
&    &
\sum_{l_{j_1}' = 1}^{l_{j_1}} \dots \sum_{l_{j_k}' = 1}^{l_{j_k}} f(l_{j_1}' \hat{e}_{j_1} + \dots + l_{j_k}' \hat{e}_{j_k})
\end{eqnarray}
where the sum includes the empty set and contributes the term $ f(0, \dots, 0) $.

So, we obtain,
\begin{widetext}
\begin{eqnarray}
&    &
\frac{d z_{\{i_1, \dots, i_k\}}}{dt} = -(\kappa_{\{i_1, \dots, i_k\}} + \bar{\kappa}(t)) z_{\{i_1, \dots, i_k\}}
+ 2 e^{-\mu/2} \sum_{l_{i_1} = 1}^{\infty} \dots \sum_{l_{i_k} = 1}^{\infty}
\sum_{\{j_1, \dots, j_n\} \subseteq \{i_1, \dots, i_k\}} 
\sum_{l_{j_1}' = 1}^{l_{j_1}} \dots \sum_{l_{j_n}' = 1}^{l_{j_n}}
\prod_{l_i \in \{j_1, \dots, j_n\}} \frac{1}{l_i'!} (\frac{\tilde{\alpha}_i \mu}{2})^{l_i'}
\times \nonumber \\
&   &
\kappa_{\sum_{i \in \{j_1, \dots, j_n\}} (l_i - l_i') \hat{e}_i + \sum_{i \in \{i_1, \dots, i_k\}/\{j_1, \dots, j_n\}} l_i \hat{e}_i} 
z_{\sum_{i \in \{j_1, \dots, j_n\}} (l_i - l_i') \hat{e}_i + \sum_{i \in \{i_1, \dots, i_k\}/\{j_1, \dots, j_n\}} l_i \hat{e}_i}
\nonumber \\
&   &
= -(\kappa_{\{i_1, \dots, i_k\}} + \bar{\kappa}(t)) z_{\{i_1, \dots, i_k\}}
+ 2 e^{-\mu/2} 
\times \nonumber \\
&   &
\sum_{\{j_1, \dots, j_n\} \subseteq \{i_1, \dots, i_k\}, \{h_1, \dots, h_m\} \equiv \{i_1, \dots, i_k\}/\{j_1, \dots, j_n\}}
\sum_{l_{h_1} = 1}^{\infty} \dots \sum_{l_{h_m} = 1}^{\infty}
\sum_{l_{j_1} = 1}^{\infty} \dots \sum_{l_{j_n} = 1}^{\infty}
\sum_{l_{j_1}' = 1}^{l_{j_1}} \dots \sum_{l_{j_n}' = 1}^{j_n}
\prod_{i \in \{j_1, \dots, j_n\}} \frac{1}{l_i'!} (\frac{\tilde{\alpha}_i \mu}{2})^{l_i'}
\times \nonumber \\
&   &
\kappa_{\sum_{i \in \{j_1, \dots, j_n\}} (l_i - l_i') \hat{e}_i + \sum_{i \in \{i_1, \dots, i_k\}/\{j_1, \dots, j_n\}} l_i \hat{e}_i} 
z_{\sum_{i \in \{j_1, \dots, j_n\}} (l_i - l_i') \hat{e}_i + \sum_{i \in \{i_1, \dots, i_k\}/\{j_1, \dots, j_n\}} l_i \hat{e}_i}
\nonumber \\
&   &
= -(\kappa_{\{i_1, \dots, i_k\}} + \bar{\kappa}(t)) z_{\{i_1, \dots, i_k\}} + 2 e^{-\mu/2} 
\times \nonumber \\
&   &
\sum_{\{j_1, \dots, j_n\} \subseteq \{i_1, \dots, i_k\}, \{h_1, \dots, h_m\} \equiv \{i_1, \dots, i_k\}/\{j_1, \dots, j_n\}}
\sum_{l_{j_1}' = 1}^{\infty} \dots \sum_{l_{j_n}' = 1}^{\infty}
\prod_{i \in \{j_1, \dots, j_n\}} \frac{1}{l_i'!} (\frac{\tilde{\alpha}_i \mu}{2})^{l_i'}
\times \nonumber \\
&   &
\sum_{l_{j_1} - l_{j_1}' = 0}^{\infty} \dots \sum_{l_{j_n} - l_{j_n}' = 0}^{\infty}
\sum_{l_{h_1} = 1}^{\infty} \dots \sum_{l_{h_m} = 1}^{\infty}
\kappa_{\sum_{i \in \{j_1, \dots, j_n\}} (l_i - l_i') \hat{e}_i + \sum_{i \in \{i_1, \dots, i_k\}/\{j_1, \dots, j_n\}} l_i \hat{e}_i} 
\times \nonumber \\
&   &
z_{\sum_{i \in \{j_1, \dots, j_n\}} (l_i - l_i') \hat{e}_i + \sum_{i \in \{i_1, \dots, i_k\}/\{j_1, \dots, j_n\}} l_i \hat{e}_i}
\nonumber \\
&   &
= -(\kappa_{\{i_1, \dots, i_k\}} + \bar{\kappa}(t)) z_{\{i_1, \dots, i_k\}}
+ 2 e^{-\mu/2} \sum_{\{j_1, \dots, j_n\} \subseteq \{i_1, \dots, i_k\}} \prod_{i \in \{j_1, \dots, j_n\}} (e^{\tilde{\alpha}_i \mu/2} - 1)
\times \nonumber \\
&   &
\sum_{\{g_1, \dots, g_p\} \subseteq \{j_1, \dots, j_n\}} \kappa_{\{g_1, \dots, g_p\} \bigcup \{i_1, \dots, i_k\}/\{j_1, \dots, j_n\}}
z_{\{g_1, \dots, g_p\} \bigcup \{i_1, \dots, i_k\}/\{j_1, \dots, j_n\}}
\end{eqnarray}
\end{widetext}

Defining $ I = \{i_1, \dots, i_k\} $, we have that the sum in the final expression may be re-expressed as,
\begin{widetext}
\begin{eqnarray}
&   &
\sum_{H \subseteq I} \sum_{G \subseteq I/H}
\prod_{i \in H} (e^{\tilde{\alpha}_i \mu/2} - 1) 
\prod_{i \in G} (e^{\tilde{\alpha}_i \mu/2} - 1)
\kappa_{H \bigcup I/(H \bigcup G)} z_{H \bigcup I/(H \bigcup G)}
\nonumber \\
&   &
= \sum_{H \subseteq I} \sum_{G \subseteq I/H}
\prod_{i \in H} (e^{\tilde{\alpha}_i \mu/2} - 1) \prod_{i \in G} (e^{\tilde{\alpha}_i \mu/2} - 1)
\kappa_{I/G} z_{I/G}
\nonumber \\
&   &
= \sum_{G \subseteq I} \prod_{i \in G} (e^{\tilde{\alpha}_i \mu/2} - 1)
\kappa_{I/G} z_{I/G}
\sum_{H \subseteq I/G} \prod_{i \in H} (e^{\tilde{\alpha}_i \mu/2} - 1)
\nonumber \\
&   &
= \sum_{G \subseteq I} e^{-\lambda (\sum_{i \in I/G} \tilde{\alpha}_i) \mu/2} 
[\prod_{i \in G} (e^{\tilde{\alpha}_i \mu/2} - 1)]
\kappa_{I/G} z_{I/G}
\nonumber \\
&   &
= e^{-(\sum_{i \in I} \tilde{\alpha}_i) \mu/2} \sum_{G \subseteq I} 
[\prod_{i \in G} (1 - e^{-\tilde{\alpha}_i \mu/2})]
\kappa_{I/G} z_{I/G}
\end{eqnarray}
\end{widetext}
and so, substituting into Eq. (B3) we obtain Eq. (22) in the main text.

Now, to derive the dynamical equations for the $ z(m_1, \dots, m_N) $, we start with Eq. (23) and take into account degeneracies, which gives us,
\begin{eqnarray}
&   &
\frac{d z(m_1, \dots, m_N)}{dt} = -(\kappa(m_1, \dots, m_N) + \bar{\kappa}(t)) z(m_1, \dots, m_N)
\nonumber \\
&   &
+ 2 e^{-(1 - m_1 \alpha_1 - \dots - m_N \alpha_N) \mu/2} {n_1 \choose m_1} \times \dots \times {n_N \choose m_N}
\times \nonumber \\
&   &
\sum_{m_1' = 0}^{m_1} \dots \sum_{m_N' = 0}^{m_N}
[\prod_{n = 1}^{N} (1 - e^{-\alpha_n \mu/2})^{m_n'}]
{m_1 \choose m_1'} \times \dots \times {m_N \choose m_N'}
\times \nonumber \\
&   &
\kappa(m_1 - m_1', \dots, m_N - m_N') \frac{z(m_1 - m_1', \dots, m_N - m_N')}{{n_1 \choose {m_1 - m_1'}} \times \dots \times {n_N \choose {m_N - m_N'}}}
\nonumber \\
\end{eqnarray}
which gives Eq. (24) after some manipulation.

\end{appendix}

\end{document}